\documentclass[namedreferences,hyperref,optionalrh]{spr-sola}
\usepackage{graphicx}        
\usepackage{color}           
\usepackage{amsmath}         
\usepackage{rotating}        
\newcommand{\RSun}{{R$_{\odot}$}}   
\newcommand{\ie}{{i.e.}}     

\newcommand{\aap}{{\it Astron. Astrophys.}}
\newcommand{\aaps}{{\it Astron. Astrophys. Suppl.}}

\newcommand{\apj}{{\it Astrophys. J.}}

\newcommand{\mnras}{{\it Mon. Not. R. Astron. Soc.}}

\newcommand{\solphys}{{\it Sol. Phys.}}

\chardef\us=`\_
\begin{document}
\begin{article}
\begin{opening}
\title{A SART-Based Iterative Inversion Methodology to Infer the Solar
  Rotation Rate from Global Helioseismic Data} 
\author[addressref=aff1,corref,email={skorzennik@cfa.harvard.edu}]{\inits{S. G.}%
\fnm{Sylvain G.}~\lnm{Korzennik}}
\author[addressref={aff2,aff3},email={adarwich@ull.edu.es}]{\inits{A.}%
\fnm{Antonio}~\lnm{Eff-Darwich}}

\address[id=aff1]{Center for Astrophysics $|$ Harvard \& Smithsonian,
  Cambridge, MA 02138, U.S.A.} 
\address[id=aff2]{Universidad de La Laguna, Tenerife, 38204, Spain}
\address[id=aff3]{Instituto de Astrof\'isica de Canarias, Tenerife, 38206,
  Spain} 
\runningauthor{Korzennik and Eff-Darwich}
\runningtitle{A SART-Based Iterative Inversion Method}
\begin{abstract}

We present a new iterative rotation inversion technique based on the
Simultaneous Algebraic Reconstruction Technique developed for image
reconstruction. We describe in detail our algorithmic implementation and
compare it to the classical inversion techniques like the Regularized Least
Squares (RLS) and the Optimally Localized Averages (OLA) methods. In our
implementation, we are able to estimate the formal uncertainty on the inferred
solution using standard error propagation, and derive the averaging kernels
without recourse to any Monte-Carlo simulation.  We present the potential of
this new technique using simulated rotational frequency splittings. We use
noiseless sets that cover the range of observed modes and associate to these
artificial splittings observational uncertainties. We also add random noise to
present the noise magnification immunity of the method. Since the technique is
iterative we also show its potential when using an apriori solution. With the
right regularization this new method can outperform our RLS implementation in
precision, scope and resolution. Since it results in very different averaging
kernels where the solution is poorly constrained, this technique infers
different values. Adding such a technique to our compendium of inversion
methods will allow us to improve the robustness of our inferences when
inverting real observations and better understand where they might be biased
and/or unreliable, as we push our techniques to maximize the diagnostic
potential of our observations.

\end{abstract}
\keywords{Solar rotation, Inverse Modeling, Helioseismology}
\end{opening}
%
%
\section{Introduction}
     \label{S-Introduction} 

The analysis of helioseismic data, carried out using various methodologies and
acquired with ground- and space-based instruments, has given us a
comprehensive insight into the rotation of the Sun's interior, its temporal
variations and the correlation of these variations with some of the
characteristics of the solar activity.

Helioseismic techniques are typically categorized into two complementary
approaches: (i) global helioseismology, which measures and interprets the
$p$-modes eigenfrequencies, $\nu_{n,\ell,m}$, and (ii) local helioseismology,
which measures and interprets the local wave field observed on the solar
surface. The most commonly employed techniques in local helioseismology are
the Fourier-Hankel decomposition, the ring-diagram analysis, the time-distance
analysis, helioseismic holography, and direct modeling \citep[see][for
  additional details]{gizon2005}.

One of the most important results derived from global helioseismic data and 
techniques is the characterization of the Sun's internal rotation.
Let us recall that the
advection of the $p$-modes by the rotation lifts the degeneracy of
eigenmodes with the same azimuthal order, $m$, within a multiplet of harmonic
degree $\ell$ and radial order $n$. The functional form of the perturbation
of the eigenfrequencies, the rotational frequency splittings, resulting from
$\Omega(r, \theta)$, the solar rotation rate as a function of radius $r$ and
co-latitude $\theta$, is a Fredholm integral equation of the first kind
\citep{hansen1977}, namely
\begin{eqnarray}\label{eq-1}
  \frac{\nu_{n,\ell,+m} -\nu_{n,\ell,-m}}{2\,m}  & = &
  \int_0^R\int_0^{\pi}K_{n,\ell,m}(r, \theta)\,\Omega(r, \theta)\,dr\,d\theta
  \\\label{eq-1b}  
  & = & \frac{\Delta \nu_{n,\ell,m}}{2\,m} \pm \epsilon_{n,\ell,m}
\end{eqnarray}
where $\Delta \nu_{n,\ell,m}$ represents the $n_{\rm obs}$ observed rotational
splittings, $\epsilon_{n,\ell,m}$ represents the associated observational errors,
and $K_{n,\ell,m}(r, \theta)$ denotes the sensitivity functions of each
splitting to the rotation rate, also known as rotational kernels.
These kernels are known functions of the solar model and the mode
eigenfunctions.

This set of $n_{\rm obs}$ integral equations, each corresponding to a distinct
frequency splitting as per Equations \ref{eq-1} and \ref{eq-1b}, defines a classical
inverse problem: these equations can be numerically inverted to infer the
radial and latitudinal variation of the rotational rate from the splittings.

Regardless of which helioseismic observations and which inversion procedures
were used, the inferred rotational rate within specific regions of the solar
interior has revealed distinctive features. These inferences have
presented inherent challenges to characterize them precisely
\citep{thompson1996, chaplin-2008, schou2008, howe-2009,howe-2016, basu2016}.
The picture that has emerged is that of a solar radiative interior that 
rotates at $\approx 431$ nHz from the base of the convection zone ($\approx
0.71$\RSun) down to $\approx 0.25$\RSun, without any significant
angular differential rotation.
This result is consistent even when using data sets from different helioseismic
instruments \citep{eff-darwich-2013}.  {The difficulty in detecting and characterizing 
with reasonably low uncertainty the lowest-degree $p$-modes makes it challenging to 
infer the rotation rate below $\approx 0.25$\RSun. One would need to reduce the 
observational errors of modes with $\ell < 5$ by a factor of approximately 50 to 
infer the general trend of the rotational rate below $\approx 0.15$\RSun 
\citep{eff-darwich2008}. Hence, when using $p$-modes alone, it is not 
possible to rule out either a fast or slow rotating core.}
The inclusion of $g$-modes, modes that are highly sensitive to the solar core,
would greatly improve inferences of the solar rotation in the deepest region
of the Sun. Some promising yet unconfirmed attempts to detect and characterize
$g$-modes have been carried out \citep{savita-2008, fossat-2017,
  scherrer-2019}.

The same pattern of differential rotation that is observed at the solar
surface is inferred within the solar convection zone ($r > 0.71$\RSun), with
the equator rotating faster than the poles. The outer $5\%$ of the solar
radius displays a sharp and intriguing increase in the rotation rate with depth.
This near surface radial gradient decreases with latitudes, and we cannot rule out 
that it may even be reversed close to the poles 
\citep[see for example,][and references therein]{Korzennik-1990, Rhodes-1990, howe-2009, barekat-2014}.
{As in the case of the core, the $p$-modes that are more sensitive to
the polar regions are the most difficult to characterize, namely the
low azimuthal order and zonal modes.  Hence, our inferences above
$\approx 75^{\circ}$ in latitude are not yet reliable.}

The transition region between the rigid rotation in the radiative zone and the
differential rotation in the convection zone is known as the
tachocline. It is characterized by a strong radial shear that might be the
source of the magnetic activity of the Sun. Hence, it is essential to study its
spatial distribution and temporal variability. The very precise position and
thickness of the tachocline and how they vary with latitude are still a
matter of debate due to the somewhat limited spatial resolution of rotation
inversions, especially in the radial direction and at at high latitudes, 
although the actual
thickness is likely to be smaller than $1\%$ of the solar radius
\citep{corbard-1998}. There are indications, although not confirmed, that some
properties of the tachocline change with time \citep{howe-2000}; what has been
confirmed is that the tachocline is prolate and thicker at higher latitudes
\citep{antia-2011}.

When a temporal mean at each depth and co-latitude is subtracted from a time
series of rotational inversions, a pattern of bands of faster and slower
rotation propagating from mid latitudes to the equator is revealed, and
penetrates down to at least $\approx 0.9$\RSun
\citep{vorontsov-2002,basu-2019} {and maybe even a little deeper \citep{howe-2005}}. 
At higher latitudes, there are bands of
faster rotation that migrate towards the poles, and penetrate the entire
convection zone. Moreover there is a {definite correspondence} between these flows and
the observed surface magnetic activity, \ie, the butterfly diagram.

In summary, global helioseismology has provided unprecedented insights into
the dynamics of the Sun's interior. But, in order to improve our
characterization of the regions closer to the solar core, the polar regions,
and the tachocline, and in order to achieve a more precise characterization of
the temporal changes of the rotation rate, it is necessary to reduce or narrow
down any biases of the data analysis \citep[see for instance][and references therein]{korzennik-2023}. 
Additionally, new
inversion methods allow us to carry out careful comparisons of inferences and thus their
reliability when using different algorithmic implementations. For these
reasons, we have developed a new methodology that is quite different from those
most commonly used for the inversion of rotational frequency splittings.

In Section 2, we summarize these most commonly used inversion techniques in
global helioseismology, whereas in Section 3, we describe a new iterative
inversion method. The performance of this iterative methodology is presented in
Section 4, using various artificial data sets, followed by our conclusions, in Section 5,
on this method niche and potential.

\section{The Basics of Helioseismic Inversion Techniques}

In practice, most inversion algorithms require discretizing the integral
relations expressed by Equations \ref{eq-1} and \ref{eq-1b} into a matrix form:
\begin{equation}\label{eq-2}
	A\, x = y  \pm \sigma_y
\end{equation}
where $y$ is the vector of observables, whose elements $y_i$ are, in the case
of rotation inversions, the rotation splittings ${\Delta\nu_{n, \ell,
    m}}/{2\,m}$, and whose dimension corresponds to the size of the observed
data set, $n_{\rm obs}$. The vector $x$ is the unknown corresponding model,
\ie, the rotation rate, that the inversion methodology is seeking to
infer. It is a vector of dimension $n_{\rm mod}$, the dimension corresponding
to the number of model grid points the solution is to be inferred on. 
The value of $n_{\rm mod}$ is
chosen to satisfy $n_{\rm mod} \ll n_{\rm obs}$, hence Equation \ref{eq-2} is
well over-constrained. The matrix $A$ corresponds to the discretized
rotational kernels and is of dimension $n_{\rm obs} \times n_{\rm mod}$. If we
use a step-wise constant discretization of the underlying model, the elements
of $A$ are given by:
\begin{equation}\label{eq-14}
a_{i,j} = \int_{r_{j-1}}^{r_j} \int_{\theta_{j-1}}^{\theta_{j}} K_{n,\ell,m}(r, \theta)
  \,dr \,d\theta ~~~~ {\rm for} ~~ i=1,n_{\rm obs} ~~{\rm and}\ \ j=1,n_{\rm mod}
\end{equation}

Since the rotation inversion is a two-dimensional problem, let us introduce
$n^{(r)}_{\rm mod}$ and $n^{(\theta)}_{\rm mod}$, the number of model grid
points in radius and co-latitude, respectively, with, of course, $n^{(r)}_{\rm
  mod} \times n^{(\theta)}_{\rm mod} = n_{\rm mod}$.
As for the observational errors, they are assumed to be
normally-distributed with zero mean and uniform variance $\sigma_y$.

Basically, two types of linear fully two-dimensional inversion techniques, plus
some modifications and adaptations, have been used in global helioseismology
to infer the internal rotation profile from the observed frequency splittings:
the Regularized Least Squares (RLS) method and the Optimally Localized
Averages (OLA) one.

The RLS approach to the inversion problem
\citep{christensen-1993,barrett-1994} is to find, essentially through a
least-squares fit, the model profile that best fits the data, subject to a
smoothness penalty term, or regularization. This is accomplished by minimizing
the following functional form:
\begin{equation}\label{eq-21}
	||A\, x - y||^2 + \alpha ||Ly||^2
\end{equation}
where an estimate of $x$ is computed by solving this set of normal
equations:
\begin{equation}\label{eq-3}
	x = (A^TA + \alpha L^TL)^{-1}A^Ty = Ty  
\end{equation}
where 
\begin{equation}\label{eq-19}
	T = (A^TA + \alpha L^TL)^{-1}A^T 
\end{equation}

The matrix $L$ in Equation \ref{eq-21} corresponds to the regularization function
that is introduced to lift the problem intrinsic singularity and in practice
to remove the oscillatory component of the solution, resulting from the
ill-conditioned nature of the problem and the effect of the noise present in
the input data set (also known as noise amplification or noise magnification).
As a result, the quality of the solution, in terms of error propagation and
damping of undesired features, depends on the choice of $L$ and the factor
$\alpha$ (also known as a Lagrange multiplier). Most used regularization
matrices, $L$, are the identity matrix, $I$, and finite difference matrices,
such as:
\begin{equation} \label{eq-l1}
 L = \frac{1}{2} \begin{bmatrix}
 1 & -1 &    &   &   &  0 \\
   &  1 & -1 &   &   &    \\
   &  . &  . & . &   &    \\
   &    &  . & . & . &    \\
 0 &    &    &   & 1 & -1 \\
\end{bmatrix}
\end{equation}
or
\begin{equation} \label{eq-l2}
L = \frac{1}{4} \begin{bmatrix}
 -1 &  2 & -1 &    &    &   &  0 \\
    & -1 &  2 & -1 &    &   &   \\
    &  . &  . &  . &    &   &   \\
    &    &  . &  . & .  &   &   \\
  0 &    &    &    & -1 & 2 & -1 \\
\end{bmatrix}
\end{equation}
By substituting $y$ in Equation \ref{eq-3} by the corresponding
expression in Equation \ref{eq-1}, the solution $x_{j}$, at the grid
point $j$ is given by:
\begin{equation}\label{eq-20}
  x_{j} = \int_0^R\int_0^{\pi} T^{(j)} K_{n,\ell,m}(r, \theta)\Omega(r,
  \theta)\,dr\,d\theta
\end{equation}
namely that the product of the $j^{th}$ row of $T$, $T^{(j)}$, by the
rotation kernel $K_{n,\ell,m}(r, \theta)$ is nothing other than the inversion
corresponding averaging kernel.
The averaging kernel is the weighting function that relates the true
underlying rotation profile, $\Omega(r_j, \theta_j)$ to the inferred profile,
$x_{j}$, at each given depth and latitude. In the ``{perfect}'' case, the
averaging kernels would simply be Dirac functions centered at $(r_j,
\theta_j)$.
Moreover, if the errors in the input data set are uncorrelated and
properly described by a normal distribution, the norm $||T||$, will
correspond to the error propagation on the solution.

In the case of the OLA approach \citep{backus1968} the most used
adaptation in helioseismology is the Subtractive Optimally Localized Averages
\citep[SOLA;][]{pijpers1994}. This inversion method looks for the
minimization, at each $j^{th}$ location where the inversion is carried out, of
the difference between the actual averaging kernels and a target kernel,
represented by a matrix $R^{(j)}$, for example a two-dimensional Gaussian or
Lorentzian function. This is done minimizing the functional form
\begin{equation}\label{eq-29}
	||A^TT^{(j)} - R^{(j)}||^2 + \alpha_{j} ||T^{(j)}||^2
\end{equation}
that leads to solving the set of normal equations:
\begin{equation}\label{eq-25}
	(AA^T + \alpha_{j} I)T^{(j)} = AR^{(j)}  
\end{equation}
\begin{equation}\label{eq-26}
	T^{(j)} = (AA^T + \alpha_{j} I)^{-1}AR^{(j)}  
\end{equation}

Both the trade-off parameter $\alpha_{j}$, \ie, a trade-off between
resolution misfit and model variance, and the radial and latitudinal
resolution of the target kernel $R^{(j)}$ must be chosen before running the
inversion. The solution of the SOLA inversions at the $j^{th}$ location is
therefore given by
\begin{equation}\label{eq-31}
	x_{j} = T^{(j)}\,y  
\end{equation}

A comparison of Equations \ref{eq-3} and \ref{eq-31} that are used to
calculate the solutions in the RLS and SOLA techniques respectively, reveals
that basically both inversion procedures form linear combinations of the data
to derive well-localized inferences of the rotation rate at different
locations within the Sun.

Let us write the matrices $T$ in both inversion methods in terms of their
singular value decomposition (SVD)
\begin{equation}\label{eq-32}
	T = U W V^T  
\end{equation}
where $W$ is a square diagonal matrix, that contains the singular values of
matrix $T$, ordered in descending order. The smallest singular values are
responsible for the amplification of the noise and any bias present in the
data \citep{christensen-1993}. Consequently, the various inversion methods can
be seen in terms of how they filter out or truncate these undesirable low
singular values. Variations in the results from different inversion
methodologies often arise from how these techniques handle these eigenvalues
and consequently how they effectively filter the uncertainties in the data.

\section{Description of the SART-Based Iterative Inversion Method}

Our inversion algorithm is based on the Landweber iteration methodology
\citep{engl1996}, which is a gradient method for the minimization of
Equation \ref{eq-2}, based on the following iteration scheme:
\begin{equation}\label{eq-22}
  x_k = x_{k-1} - \beta\, A^T (A\,x_{k-1} - y)  
\end{equation}

The iteration scheme in Equation \ref{eq-22} is for $0 < \beta < 2/||A||^2$ 
and is none
other than a linear regularization method for the problem described in
Equation \ref{eq-2}, as long as the iteration is truncated at some finite index
$k_{\rm max}$.

In a more general framework \citep{sor2017}, Equation \ref{eq-22} can be
written as:
\begin{equation}\label{eq-23}
  x_k = x_{k-1} - \beta\, Q^{-1} (A\,x_{k-1} - y)  
\end{equation}

A family of different iterative reconstruction algorithms has been
implemented depending on the choice of the matrix $Q$ \citep{xu1993,mac1990}.
We opted to use the prescription for $Q$ given by the Simultaneous Algebraic
Reconstruction Technique (SART). SART was originally implemented to solve
linear systems in image reconstruction \citep{anderson-kak1984,sluis1990}. It
has since been proposed as an inversion technique for the solar internal
rotation rate by \citet{eff-darwich2010}.

In our implementation of the SART algorithm, the solution, $x_k$, after $k$ iterations is
given by
\begin{equation}\label{eq-4}
   x_k = x_{k-1} + \beta\, P (y - A\,x_{k-1}) - \beta \alpha L^TL x_{k-1} 
\end{equation}
where the matrix $P$ is given by
\begin{equation}\label{eq-5}
	P = B^{-1}A^TC^{-1} 
\end{equation}

The SART algorithm calculates the diagonal matrices $B$ and $C$ from the
summation of columns and rows of the matrix $A$, namely
\begin{equation}\label{eq-13}
	B_{j,j} = \sum_{i=1}^{n_{\rm mod}} A_{i,j}/{\sigma_j}^b
\end{equation}
and
\begin{equation}\label{eq-16}
	C_{i,i} = \sum_{j=1}^{n_{\rm obs}} A_{i,j}/{\sigma_j}
\end{equation}

The matrix $B$ normalizes the iteration relative to the weighting given by
each grid point to the data samples. The matrix $C$ takes into account the
weight of the samples of the data set. Both $B$ and $C$ are normalized by the
uncertainties of the observables, hence, the iterative process takes into
account the error distribution of the data. We added an extra parameter $b$ to
control the weight of the data errors in the computation of $B$, whose default
value is 1, but can be, in practice, adjusted to values between 0.75 and 1.25 
to modulate the effect of the uncertainties.

The matrix $L$ in Equation \ref{eq-4}, corresponds to that defined in Equations
\ref{eq-l1} or \ref{eq-l2}. However, it can be written more
generally as:
\begin{equation} \label{eq-l3}
L = \frac{1}{2n_{\rm nz}} \begin{bmatrix}
 l_{1,1} & l_{1,2} & l_{1,3} &         &           &           & l_{1,n_{\rm mod}} \\
         & l_{2,2} & l_{2,3} & l_{2,4} &           &           &         \\
         &     .   &     .   &     .   &           &           &         \\
         &         &     .   &     .   &   .       &           &         \\
       0 &         &         &         & l_{n_{\rm mod},n_{\rm mod}-2} & l_{n_{\rm mod},n_{\rm mod}-1} & l_{n_{\rm mod},n_{\rm mod}} \\
\end{bmatrix}
\end{equation}
for $n_{\rm nz} > 1$.
Each row $j$ ($j =1, n_{\rm mod}$) of the matrix $L$ in Equation \ref{eq-l3}
represents how the solution around each model grid point $p_i$ is
smoothed or regularized, where $p_j \equiv (r_{i'}, \theta_{j'})$, and $i'
= 1, n^{(r)}_{\rm mod}$ and $j' = 1, n^{(\theta)}_{\rm mod}$.

If for each model grid point, $p_j$, we set a regularization interval defined by
a width in radius, $W^{(r)}_{j}$ and a width in co-latitude,
$W^{(\theta)}_{j}$, the corresponding row $j$ of $L$ is given by:
\begin{eqnarray} \label{eq-l5}
  l_{j,s} = -1 & {\rm if}  & r(p_j)      - W^{(r)}_{j}     < r_s      \le r(p_j)      + W^{(r)}_{j} \\
              & {\rm and} & \theta(p_j) - W^{(\theta)}_{j} < \theta_s \le \theta(p_j) + W^{(\theta)}_{j}
  \ \ \ {\rm for}\ s =1, n_{\rm mod} \nonumber \\
  l_{j,s} = 0 & {\rm otherwise}
  \nonumber
\end{eqnarray}
where we define $n_{\rm nz}$ as the number of non-zero values of the row $l_{j,s}$, and
where $r(p_j)$ represents the radius at the model grid point $p_j$, while
$\theta(p_j)$ represents the co-latitude at the model grid point $p_j$.

Moreover, we have:
\begin{equation} \label{eq-l6}
  l_{j,s} = n_{\rm nz} \ \ \ \ {\rm if} \ \ r(p_j) = r_s  \ \ {\rm and}
  \ \ \theta(p_j) = \theta_s \ \ \ {\rm for}\ s=1,n_{\rm mod}
\end{equation}

The local regularization intervals, $W^{(r)}_{j}$ and $W^{(\theta)}_{j}$, are
further parametrized in terms of depth and latitude as follows:
\begin{equation} \label{eq-17}
  W = W_o \times (1 + f_r (1-r/R_{\odot})^{\gamma_r}) \times (1 + f_\theta (1-\theta/(\pi/2))^{\gamma_\theta}
\end{equation}
where $W_o$, $f_r$, $f_\theta$, $\gamma_\theta$ and $\gamma_r$ are
adjustable parameters, with one set for determining $W^{(r)}$ and one set
for $W^{(\theta)}$.

The independent radial and latitudinal regularization intervals defined in
Equations \ref{eq-l3}  to  \ref{eq-17} allow us to study in detail the impact of
different regularization schemes in the inversion procedure. On the one hand, the
spatial distribution of the inversion model grid is an adjustable parameter,
namely the number of grid points in both directions (\ie, the values of
$n^{(r)}_{\rm mod}$ and $n^{(\theta)}_{\rm mod}$).
On the other hand, the
model grid point spacing does not need to be uniform and can be adjusted to
correspond to the inversion resolution, a property that is known a posteriori
by examining the averaging kernels and depends mostly on the extend of the
data set (\ie, which modes are included) and the uncertainties.

After reordering terms, Equation \ref{eq-4} becomes:
\begin{equation}\label{eq-7}
	x_k = (I - \beta PA - \beta \alpha L^TL ) x_{k-1} + \beta Py 
\end{equation}
and since $x_{k-1}$ can be written in terms of $x_{k-2}$, we have
\begin{equation}\label{eq-8}
	x_k = (I - \beta PA - \beta \alpha L^TL)[(I - \beta PA - \beta \alpha L^TL) x_{k-2} + \beta Py] + \beta Py 
\end{equation}

Again, after reordering terms, we can write:
\begin{equation}\label{eq-9}
	x_k = (I - \beta PA - \beta \alpha L^TL) x_{k-2} + [(I - \beta PA - \beta \alpha L^TL) + I]\beta Py
\end{equation}

This process can be repeated for each iteration, namely:
\begin{equation}\label{eq-10}
\begin{aligned}
	x_k = (I - \beta PA - \beta \alpha L^TL )^{k} x_{0} + [(I - \beta PA - \beta \alpha L^TL)^{k-1} + \\ 
	(I - \beta PA - \beta \alpha L^TL)^{k-2}  + (I - \beta PA - \beta \alpha L^TL)^{k-3} + ... + I]\beta Py
\end{aligned}
\end{equation}
leading to:
\begin{equation}\label{eq-11}
	x_k=(I - \beta PA - \beta \alpha L^TL)^{k} x_{0} + \sum_{i=0}^{k-1}(I - \beta PA - \beta \alpha L^TL)^{i}\beta Py
\end{equation}
that we rewrite, by introducing a matrix $T$, as
\begin{equation}\label{eq-12}
	x_k = (I - \beta PA - \beta \alpha L^TL)^{k} x_{0} + Ty 
\end{equation}
where $T$ is given by
\begin{equation}\label{eq-18}
	T = \sum_{i=0}^{k-1}(I - \beta PA - \beta \alpha L^TL )^{i}\beta P
\end{equation}

 The solution at iteration $k$ is therefore calculated in terms of an
initial estimate, $x_0$, and a linear combination of the observables,
\ie, the data vector, $y$, as per the matrix $T$.
Note that $x_0$ can in principle be anything we want, and while it is often
just an apriori estimate of the mean of the solution, it can be used to
incorporate prior information on the solution.

For any matrix $G$ that has all of its eigenvalues with absolute
values smaller than 1, the following relationship holds
\begin{equation}\label{eq-38}
  G^{-1} = \sum_{i=0}^{\infty}(1-G)^i  
\end{equation}

By comparing Equation \ref{eq-19} and the left hand side of Equation \ref{eq-38}, we
can write
\begin{equation}\label{eq-35}
	G = (A^TA + \alpha L^TL)  
\end{equation}
whereas from the comparison of Equation \ref{eq-18} and the right hand side
of Equation \ref{eq-38} we  find the following approximation
\begin{equation}\label{eq-36}
	(1-G)^i \approx (I - \beta PA - \beta \alpha L^TL )^{i}  
\end{equation}
hence, the inversion matrix for the RLS method is in fact an approximation of
the truncated power expansion, up to $k$, defined by our SART-based iterative
technique.

Unlike previous attempts to calculate the error propagation and resolution of
iterative helioseismic inversions using Monte-Carlo simulations
\citep{sudnik2009}, our method gives us an explicit formulation to compute
both the error propagation and the averaging kernels by using the matrix $T$.

The choice of the optimal inversion parameters is dictated by finding the best
trade-off between the error propagation $||T||$ and the goodness of the
solution in a least squares sense, $||A\,x-y||$. In this SART-based iterative
inversion methodology, the solution depends on three primary parameters:
$\alpha$, $\beta$ and $k_{\rm max}$ the number of iterations, although besides
these, the results will also depend on the choice of the regularization matrix
$L$ and the inversion model grid. {In practice, we used in all cases 1500 iterations, 
a number that was determined to correspond to full convergence.}

\section{Potential of the SART-Based Iterative Inversion Methodology}

The potential of this iterative technique was carefully ascertained using
artificial data. To this effect, we used realistic rotational profiles, namely
profiles that include a tachocline, with a solid body and differential
rotation below and above it, but also other features placed either at the
center or closer to the surface and at high latitudes. A precise forward
computation, using rotational kernels derived from a standard model
\citep[model `$S$';][]{christensen-1996}, was carried out to produce sets of
rotational splittings that cover the range of modes that can be precisely
measured with modern helioseismic techniques, namely the $(n,\ell,m)$ set of
modes resulting from fitting a 6.3 years long time series of the Helioseismic
and Magnetic Imager (HMI) observations \citep[\ie, 2304 day long or $32 \times
  72$ days, see][and references therein]{korzennik-2023}.

These models and computations were devised and carried out in the framework
of an ongoing large collaboration to compare modern rotation inversion
techniques\footnote{These models were devised by Dr.\ A.\ Kosovichev, 
  {and already used in a ``{hare and hounds}'' exercise back in 1998, see Appendix A of \cite{schou-1998}},
  while the ongoing collaboration was initiated by the late Dr.\ M.\ Thompson and is
  now led by Dr.\ J.\ Christensen-Dalsgaard and should be eventually published
  in Christensen-Dalsgaard {\em et al.}, (in preparation).}. We reproduced
these forward computations with our own numerical implementation down to
numerical noise and used the analytical representation of one model to produce
somewhat similar models for additional forward computations of artificial data
sets.

The observational uncertainties of fitting a 6.3 year long time series were
used to associate an uncertainty to each artificial rotational splitting,
since even when inverting a noiseless artificial data set, any inverse method
needs to scale the problem by the uncertainties.  While the overall scaling of
these uncertainties is arbitrary when inverting noiseless splittings, using
uniform uncertainties would attribute more diagnostic potential to the low
degree modes. 

In addition to noiseless artificial splittings, sets of artificial splittings with
random Gaussian noise with zero mean and a standard deviation equal to each mode
uncertainty were also computed and inverted. 

\begin{figure}[!h]
\begin{center}
\includegraphics[width=0.975\textwidth]{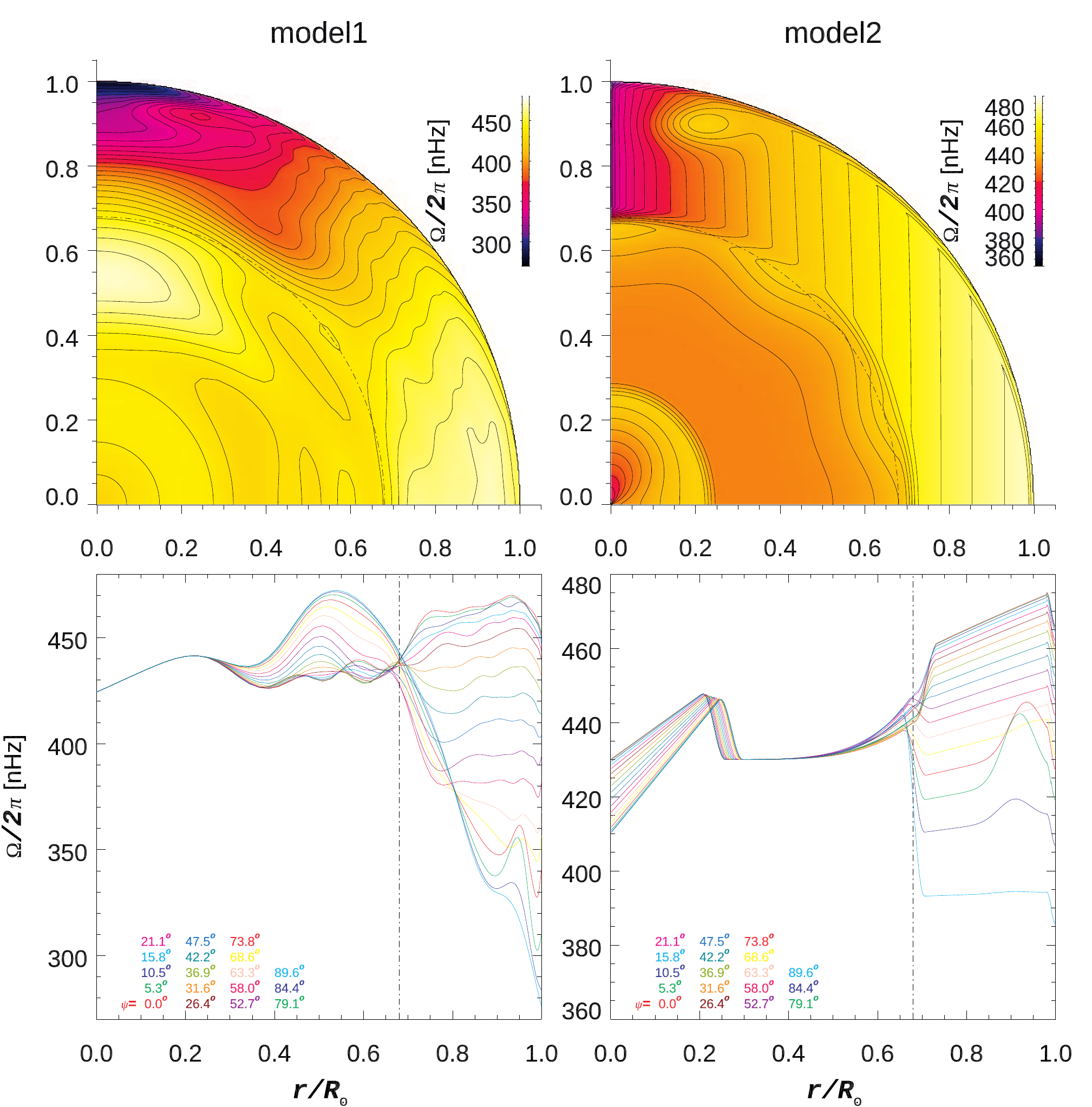} 
\end{center}
\caption{Speculative rotation profiles used to produce artificial rotational
  splittings. The two panels in the top row shows them in Cartesian
  coordinates, while the two panels in the bottom row show cuts versus radius
  at selected latitudes. The two panels on the left correspond to model 1, a
  model that includes a feature at high latitudes, a quasi-periodic
  modulation in the convection zone and a wide tachocline, while the two
  panels on the right correspond to model 2, a model that includes a high
  latitude ``jet'', a narrow tachocline with some modulation with latitude and
  a ``feature'' in the inner 30\% radius. 
  {In both cases a near-surface shear layer was included.}}\label{fig:rotProfiles}
\end{figure}

\begin{figure}
\begin{center}
\includegraphics[width=0.975\textwidth]{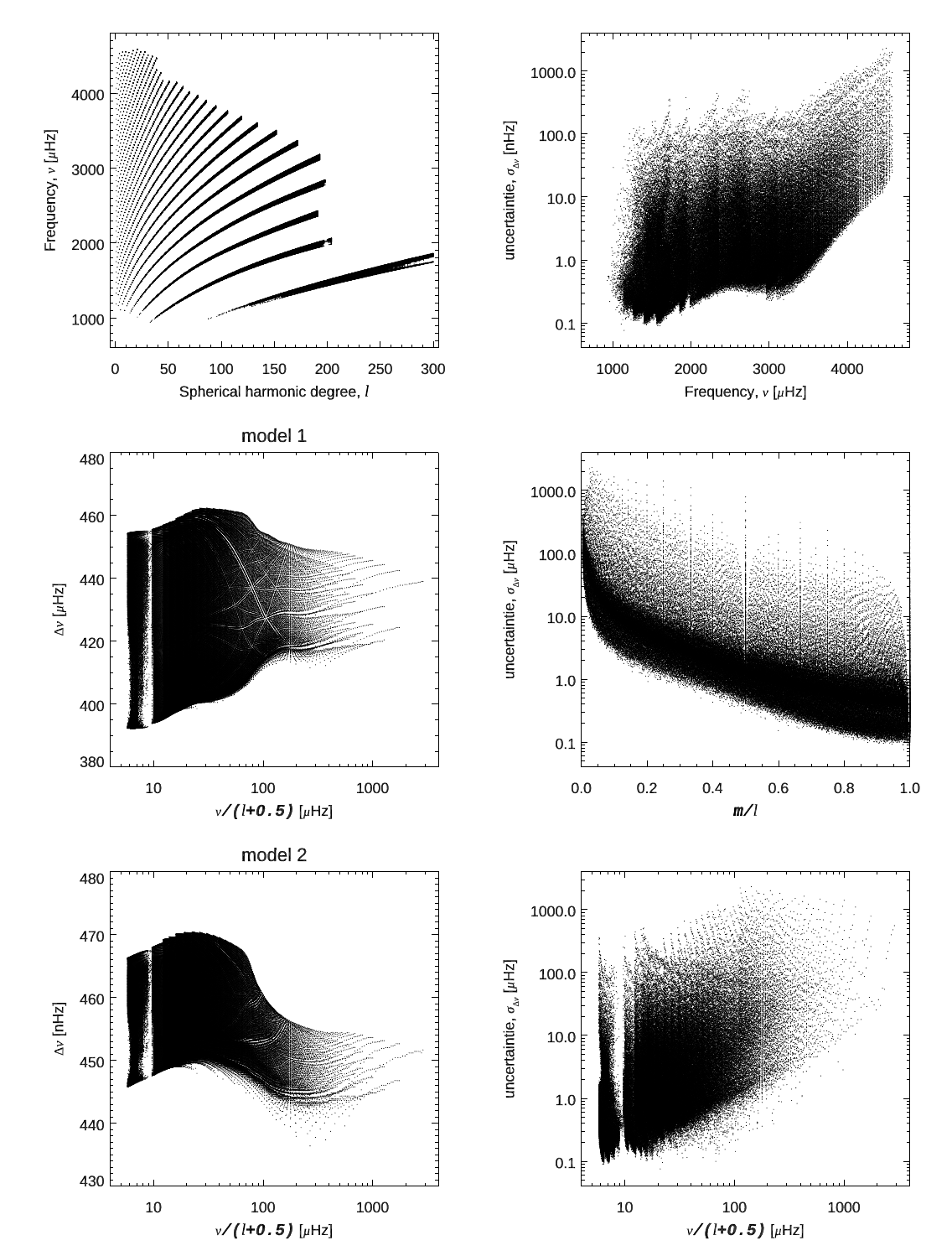} 
\end{center}
\caption{Artificial data sets: the upper left panel shows the coverage of the
  data set in the $\ell$--$\nu$ space, while the middle and bottom left panels
  show the rotational splittings for model 1 and model 2 respectively, as a
  function of $\nu/(\ell+0.5)$, a proxy for the inner turning points of
  the modes.
  The panels on the right show the uncertainties plotted versus the mode
  frequency, the $m/\ell$ ratio and $\nu/(\ell+0.5)$, in the top to bottom
  panels, respectively. }\label{fig:dataSet}
\end{figure}

Figure \ref{fig:rotProfiles} shows two rotation profiles used to generate
artificial rotational splittings, while Figure \ref{fig:dataSet} shows the
actual mode set in the $\ell$--$\nu$ space, the corresponding noiseless
splittings for each mode, and the distribution of the uncertainties.

\subsection{Inversion of Noiseless Artificial Data}

\begin{figure}
\begin{center}
\includegraphics[width=0.975\textwidth]{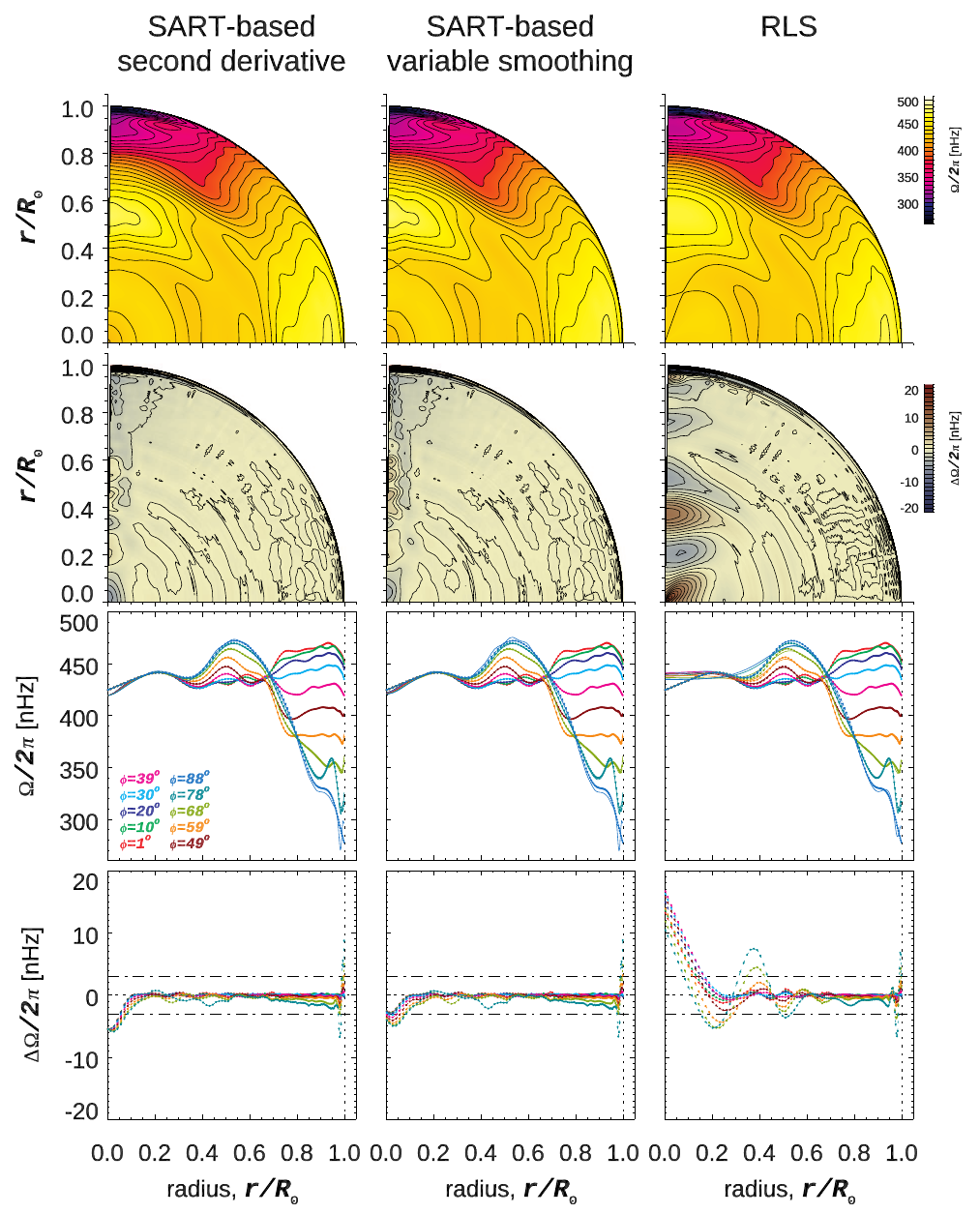} \\ 
\end{center}
\caption{Inverted solutions, based on noiseless artificial splittings,
  corresponding to model 1.
  The panels show the inferred solution and the difference between that
  inferred solution and the actual rotation model, in Cartesian coordinates in
  the top two rows or as cuts versus radius at a set of latitudes in the
  bottom two rows.
  The panels in the left column correspond to the SART-based inversion using a
  second derivative smoothing, the panels in the middle column correspond to the
  SART-based inversion using a variable smoothing used for regularization, while
  the panels in the rightmost column show the solution when using a conventional RLS
  inversion methodology.}\label{fig:mod1_noerr}
\end{figure}

\begin{figure}
\begin{center}
\includegraphics[width=0.975\textwidth]{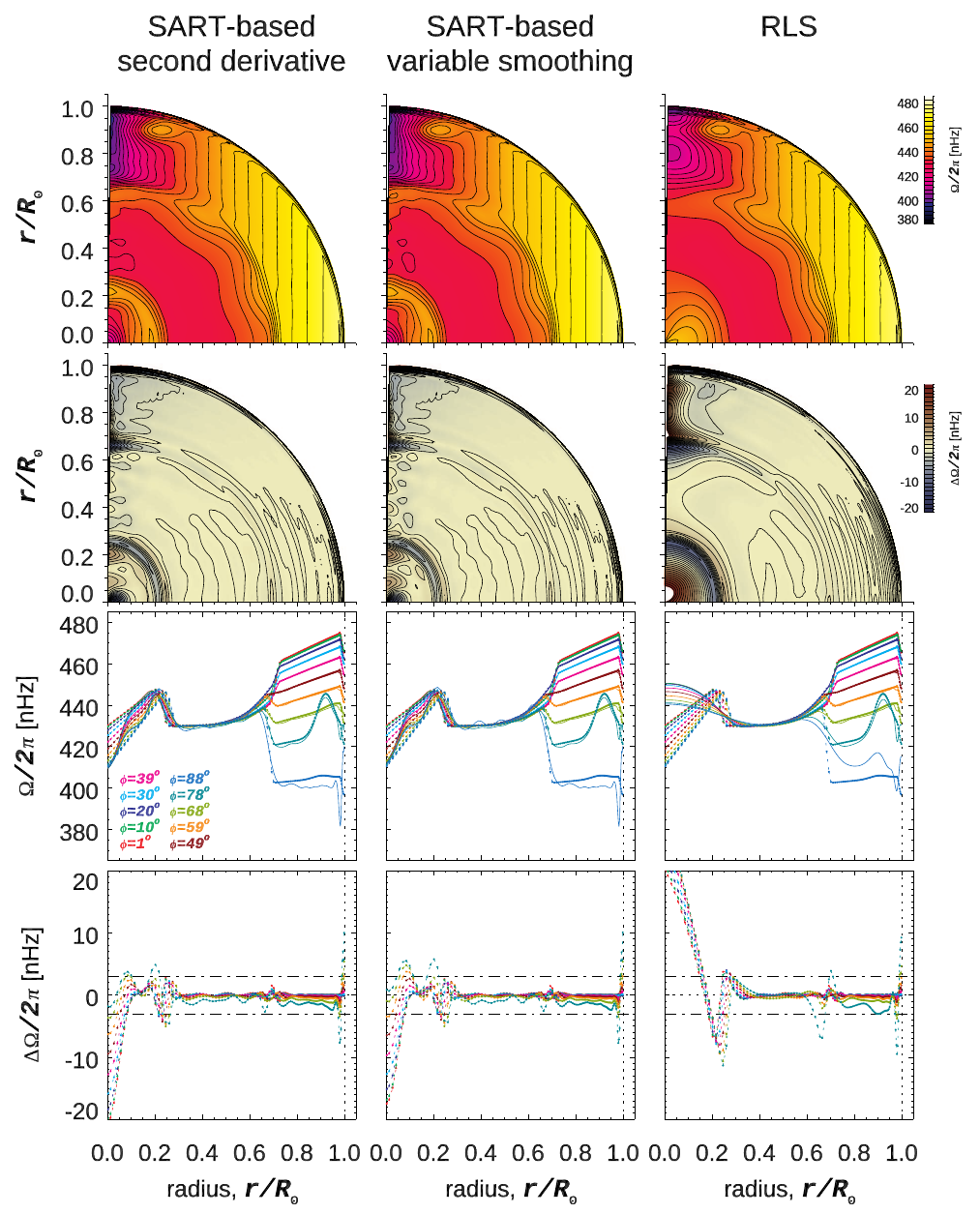} 
\end{center}
\caption{Inverted solution, based on noiseless artificial splittings,
  corresponding to model 2.
  The panels show the inferred solution and the difference between that
  inferred solution and the actual rotation model, in Cartesian coordinates in
  the top two rows or as cuts versus radius at a set of latitudes in the
  bottom two rows.
  The panels in the left column correspond to the SART-based inversion using a
  second derivative smoothing, the panels in the middle column correspond to the
  SART-based inversion using a variable smoothing used for regularization, while
  the panels in the rightmost column show the solution when using a conventional RLS
  inversion methodology.}\label{fig:mod2_noerr}
\end{figure}

Figures~\ref{fig:mod1_noerr} and \ref{fig:mod2_noerr} show the inferred
rotational profiles when inverting noiseless artificial splittings for each
model respectively. Each figure presents the inferred rotation rate using the
SART-based inversion method when using two different regularizations, namely a
second derivative smoothing or a variable smoothing \footnote{ 
  parameters of the variable regularization are $(W_o, f_r, f_\theta,
  \gamma_\theta, \gamma_r)$ = (1E-2, 99, 1, 3, 3) for $W^{(r)}$ and (3.6E-2,
  0, 1, 1, 3) for $W^{(\theta)}$.} that is a function of depth and
latitude. They also show the inferred profile when using a conventional RLS
inversion methodology {\citep{eff-darwich1997}}.  For all three inversion
methods most of the ``{features}'' of the models are recovered, although
different systematic biases can be seen.

Most notably, for model 1, the values in the inner 25\% radius show
systematic errors larger than elsewhere and the two SART-based inferences are
overall more accurate than the RLS one.
The same can be seen for model 2, although the improved ``{sharpness}'' of
the SART-based inferences is more obvious in this case since the model 2
rotation rate has such sharp and distinct features. Again, the SART-based
inferences perform better closer to the center than the RLS, but still show
systematic errors in the inner 25\% radius of a different nature than the
errors in the RLS inference.
For both models, the SART-based inferences are more accurate than the
RLS ones in the outer 75\% radius and at high latitudes. This is summarized
in Table~\ref{tab:noerr} where the RMS of the differences with respect to the
input model profile are tabulated.

\begin{figure}\begin{center}
    \includegraphics[width=0.975\textwidth]{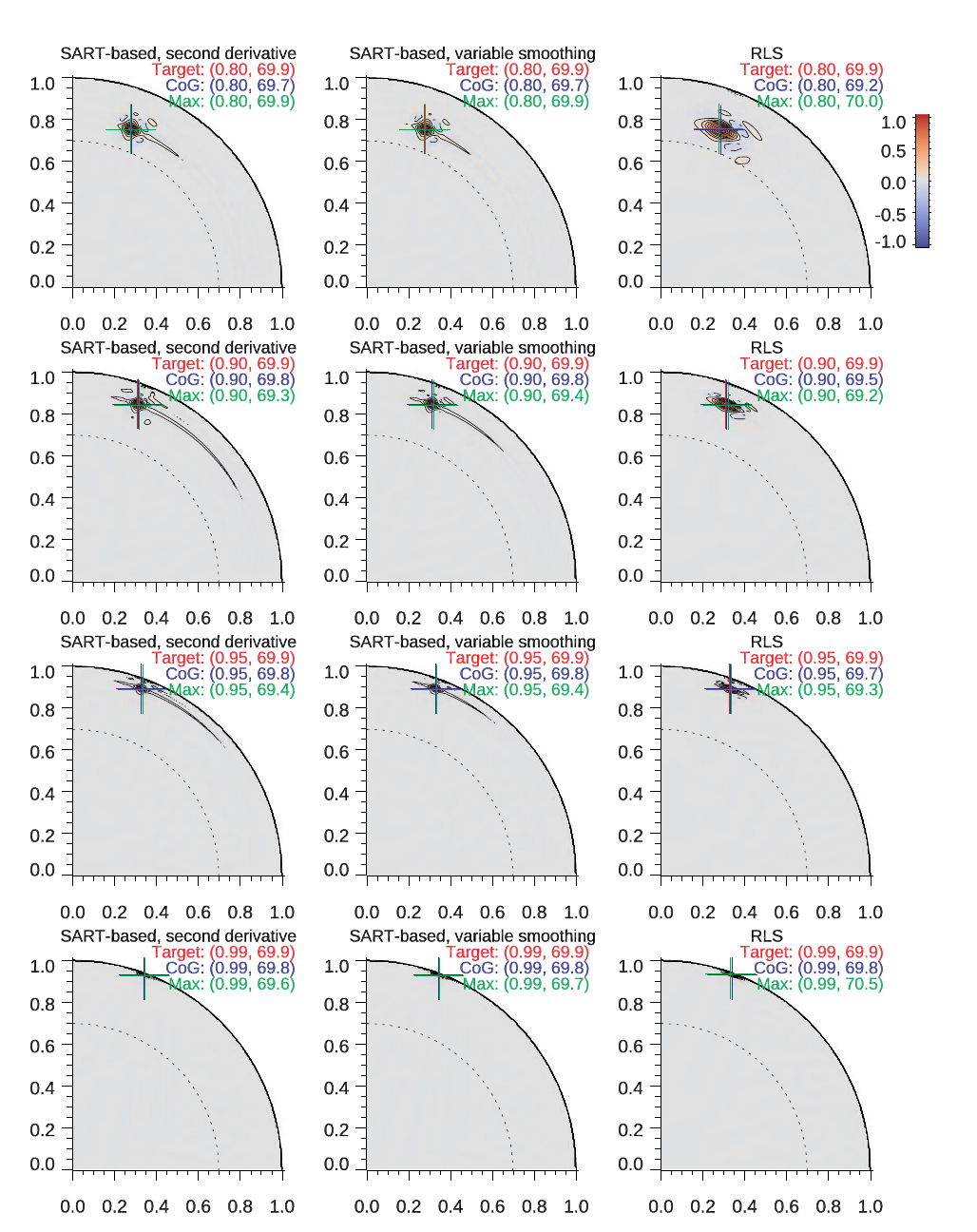} 
  \end{center}
  \caption{Averaging kernels for solutions shown in Figures \ref{fig:mod1_noerr} and
    \ref{fig:mod2_noerr}, when using noiseless artificial splittings.
    The panels in each row show them at various depths near the surface
    ($r/R_\odot=0.8, 0.9, 0.95, 0.99$) at a high target latitude, namely
    $\phi=70^{\circ}$. 
    The panels on the left correspond to the SART-based method when using a
    second derivative smoothing, the panels in the middle to the SART-based
    method when using a variable smoothing for regularization, while the panels
    on the right correspond to using a RLS inversion
    methodology.
    The position of the target location, the center of gravity, and the maximum
    of the kernels are color coded, marked as crosses and labeled as $(r,
    \phi)$.}\label{fig:avgkern-noerr-l70s}  
\end{figure}

\begin{figure}\begin{center}
    \includegraphics[width=0.975\textwidth]{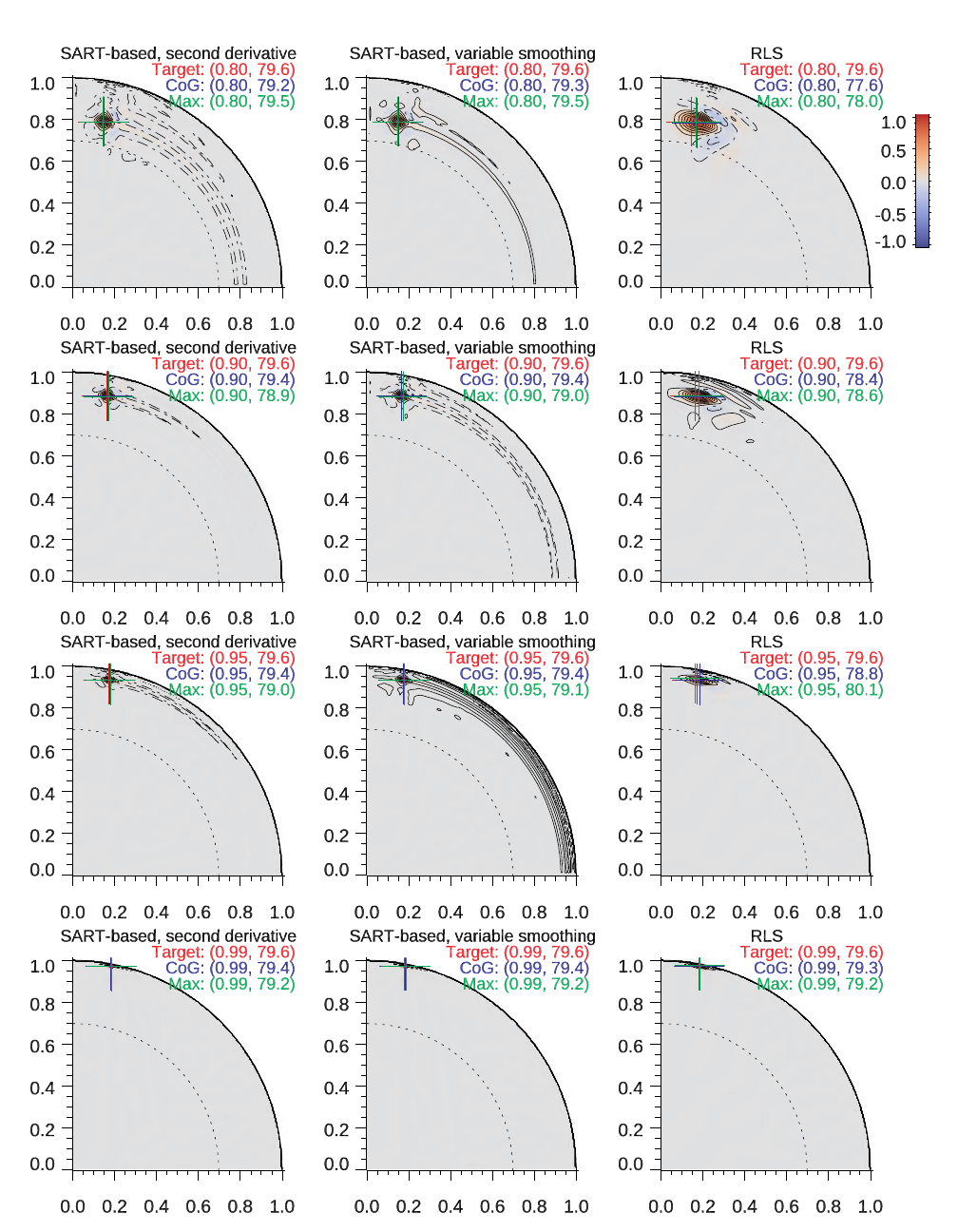} 
\end{center}
\caption{Averaging kernels for solutions shown in Figures \ref{fig:mod1_noerr} and
  \ref{fig:mod2_noerr}, when using noiseless artificial splittings.
  The panels in each row show them at various depths near the surface
  ($r/R_\odot=0.8, 0.9, 0.95, 0.99$) at an even higher target latitude, namely
  $\phi=80^{\circ}$.
  The panels on the left correspond to the SART-based method when using a
  second derivative smoothing, the panels in the middle to the SART-based
  method when using a variable smoothing for regularization, while the panels
  on the right correspond to using a RLS inversion
  methodology.
  The position of the target location, the center of gravity, and the maximum
  of the kernels are color coded, marked as crosses and labeled as $(r,
  \phi)$.}\label{fig:avgkern-noerr-l80s}  
\end{figure}

\begin{figure}
  \begin{center}
    \includegraphics[width=0.975\textwidth]{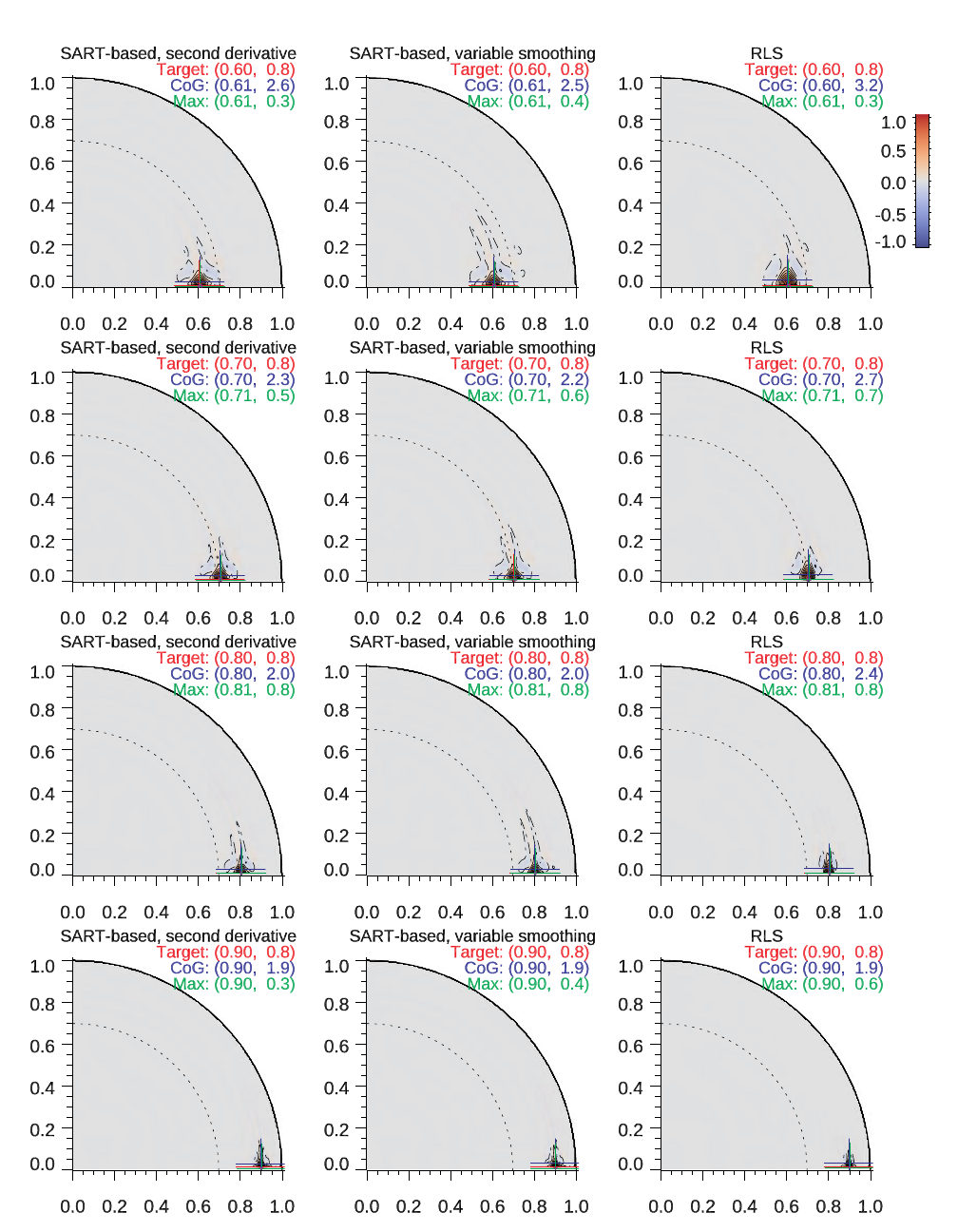} 
  \end{center}
  \caption{ Averaging kernels for solutions shown in Figures \ref{fig:mod1_noerr} and
    \ref{fig:mod2_noerr}, when using noiseless artificial splittings. 
    The panels in each row show them at various depths in the outer 40\% 
    ($r/R_\odot=0.6, 0.7, 0.8, 0.9$) at the equatorial target latitude, namely
    $\phi=0^{\circ}$.
    The panels on the top row correspond to the SART-based method when using a
    second derivative smoothing, the panels in the middle row to the SART-based
    method when using a variable smoothing for regularization, while the panels
    in the bottom row correspond to using a RLS inversion
    methodology.
    The position of the target location, the center of gravity, and the maximum
    of the kernels are color coded, marked as crosses and labeled as $(r,
    \phi)$.}\label{fig:avgkern-noerr-l00m}
\end{figure}

\begin{figure}
  \begin{center}
    \includegraphics[width=0.975\textwidth]{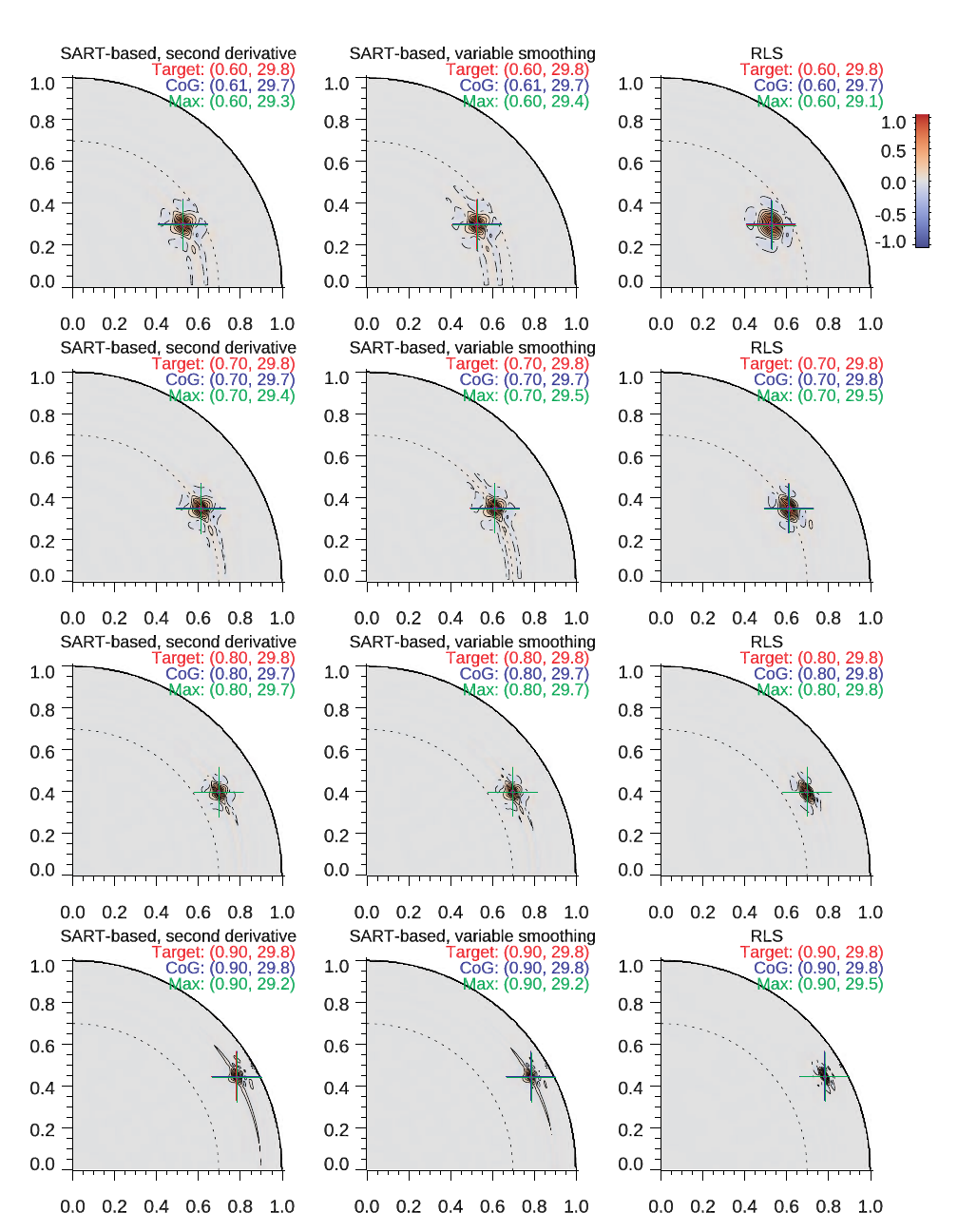} 
  \end{center}
  \caption{Averaging kernels for solutions shown in Figures \ref{fig:mod1_noerr} and
    \ref{fig:mod2_noerr}, when using noiseless artificial splittings. 
    The panels in each row show them at various depths in the outer 40\% 
    ($r/R_\odot=0.6, 0.7, 0.8, 0.9$) at a low target latitude, namely
    $\phi=30^{\circ}$.
    The panels on the left correspond to the SART-based method when using a
    second derivative smoothing, the panels in the middle to the SART-based
    method when using a variable smoothing for regularization, while the panels
    on the right correspond to using a RLS inversion
    methodology.
    The position of the target location, the center of gravity, and the maximum
    of the kernels are color coded, marked as crosses and labeled as $(r,
    \phi)$.}\label{fig:avgkern-noerr-l30m}
\end{figure}

\begin{figure}
  \begin{center}
    \includegraphics[width=0.975\textwidth]{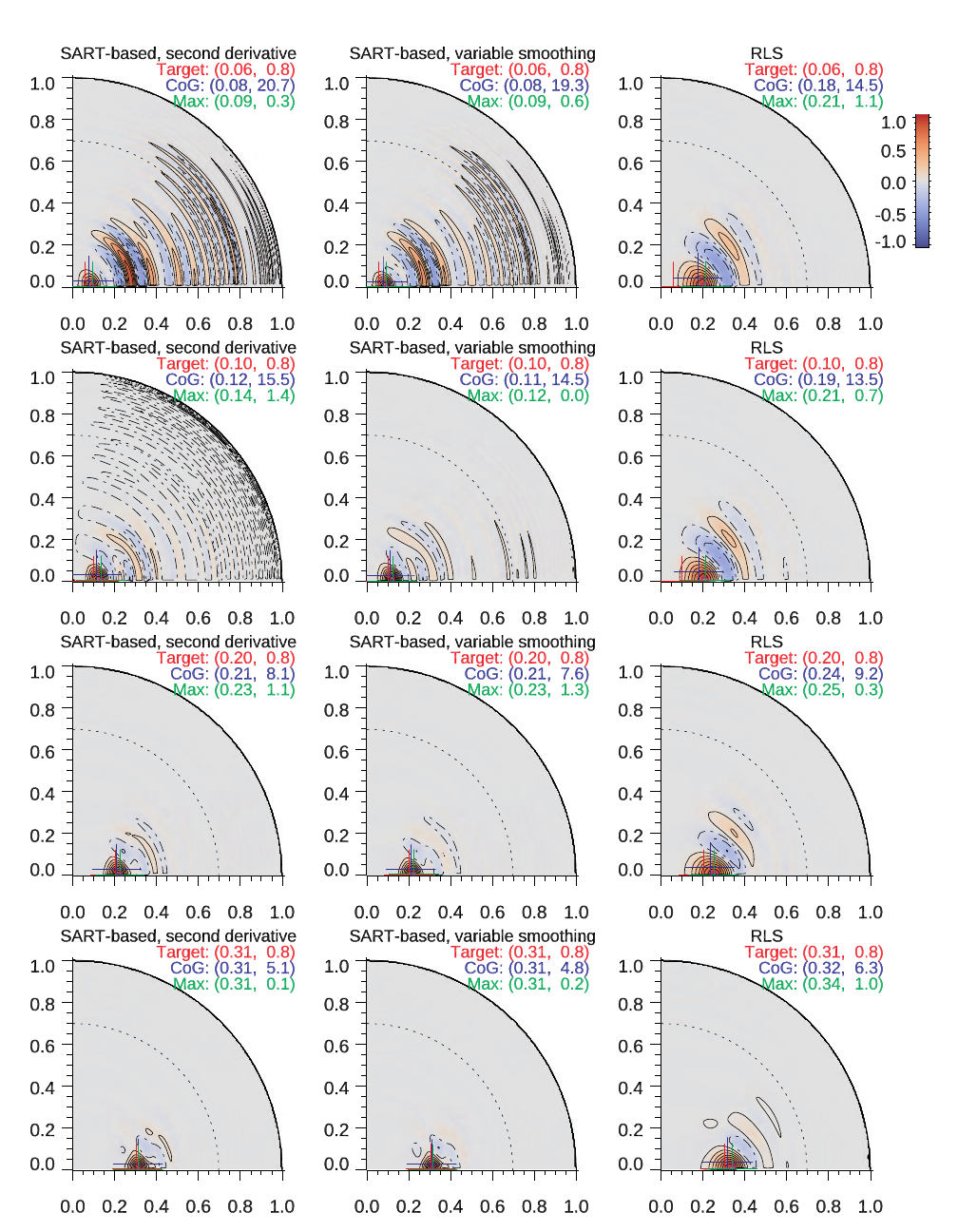} 
  \end{center}
  \caption{Averaging kernels for solutions shown in Figures \ref{fig:mod1_noerr} and
    \ref{fig:mod2_noerr}, when using noiseless artificial splittings.
    The panels in each row show them at various depths in the inner 30\% 
    ($r/R_\odot=0.06, 0.1, 0.2, 0.3$) at the equatorial target latitude, namely
    $\phi=0$. 
    The panels on the left correspond to the SART-based method when using a
    second derivative smoothing, the panels in the middle to the SART-based
    method when using a variable smoothing for regularization, while the panels
    on the right correspond to using a RLS inversion
    methodology.
    The position of the target location, the center of gravity, and the maximum
    of the kernels are color coded, marked as crosses and labeled as $(r,
    \phi)$.}\label{fig:avgkern-noerr-l00c}
\end{figure}

\begin{figure}
  \begin{center}
    \includegraphics[width=0.975\textwidth]{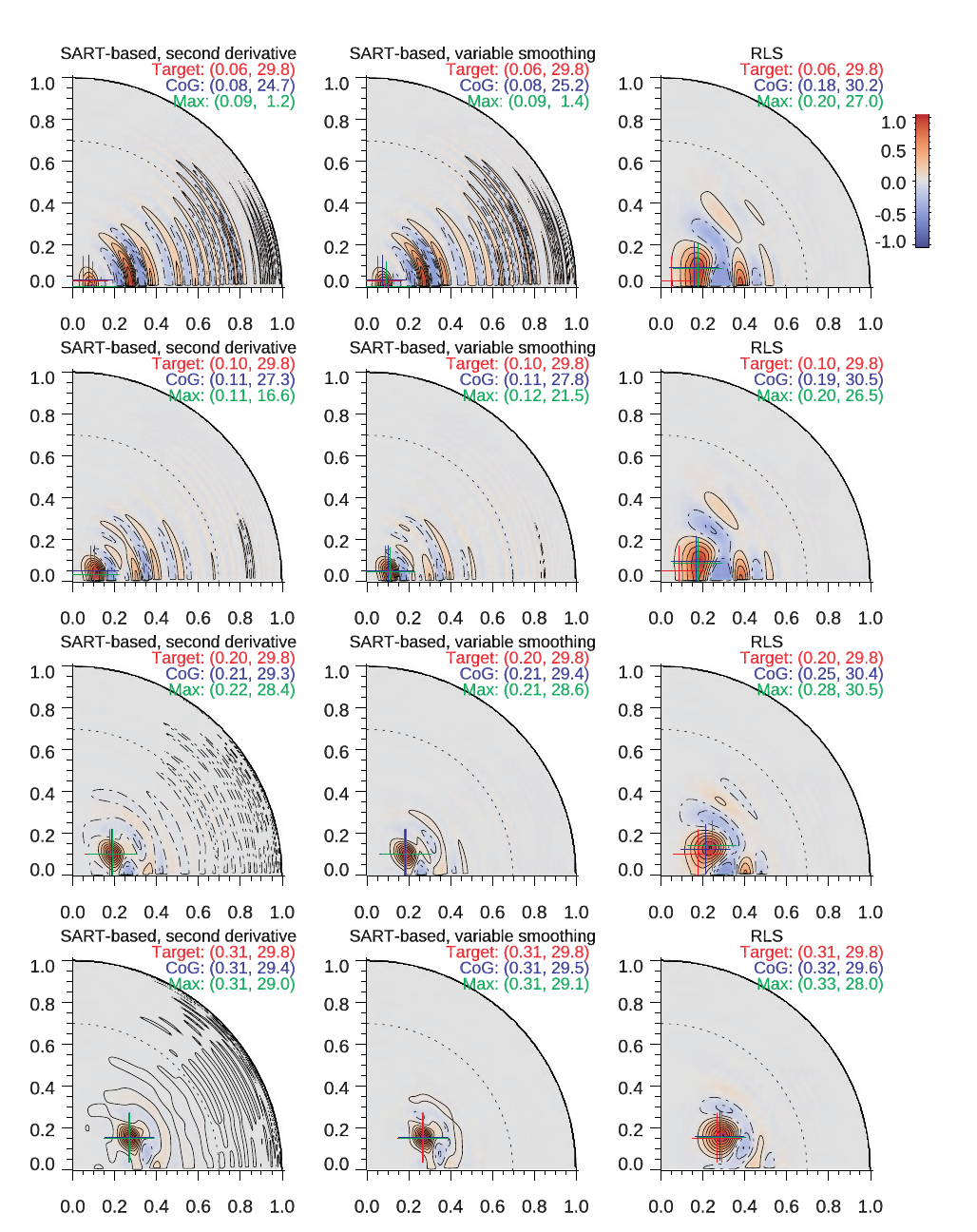} 
  \end{center}
  \caption{Averaging kernels for solutions shown in Figures \ref{fig:mod1_noerr} and
    \ref{fig:mod2_noerr}, when using noiseless artificial splittings.
    The panels in each row show them at various depths in the inner 30\% 
    ($r/R_\odot=0.06, 0.1, 0.2, 0.3$) at a low target latitudes, namely
    $\phi=30^{\circ}$. 
    The panels on the left correspond to the SART-based method when using a
    second derivative smoothing, the panels in the middle to the SART-based
    method when using a variable smoothing for regularization, while the panels
    on the right correspond to using a RLS inversion
    methodology.
    The position of the target location, the center of gravity, and the maximum
    of the kernels are color coded, marked as crosses and labeled as $(r,
    \phi)$.}\label{fig:avgkern-noerr-l30c}
\end{figure}

The corresponding averaging kernels, for selected target locations are shown
in Figures \ref{fig:avgkern-noerr-l70s} to \ref{fig:avgkern-noerr-l30c} for
target locations near the surface, at mid depth or near the center 
and at low or high latitudes.
Let us point out that averaging kernels, for all else equal, are independent
of the values of the splittings (\ie, of rotation profile), since it is the
extent of the mode set, the splittings uncertainties and the trade-off value
that matters.

For most target locations the averaging kernels resulting from each inversion
technique and type of regularization are somewhat dissimilar, with the largest
contrast for the kernels corresponding to the deeper target locations. It is
worth noting that the properties of the averaging kernels resulting from using
the SART-based methodology
are such that their centers of gravity are almost always very
close to the target location, although the center of gravity of the main peak
of that kernel can be quite displaced from that target location. This is quite
different from the similar property of the averaging kernels resulting from using 
the RLS
methodology, where the overall center of gravity of the kernels can be quite
different from the target location when the kernel is not well localized 
at that target location. One can also see that the kernels
resulting from the SART-based methodology have, when the kernel is well
localized, a more narrow main peak than those resulting from the RLS
methodology, hence the inverted profiles are sharper, but also show more
extended ``{wings}''. In contrast, when the kernel is poorly-localized, the
kernels for the SART-based methodology have very extended ``{wings}'', while the
kernels for the RLS methodology are more compact, but their center of gravity
is quite offset from the target location (see
Figure \ref{fig:avgkern-noerr-l00c}).

\subsection{Inversion of Artificial Data with Random Noise}

\begin{figure}
  \begin{center}
    \includegraphics[width=0.975\textwidth]{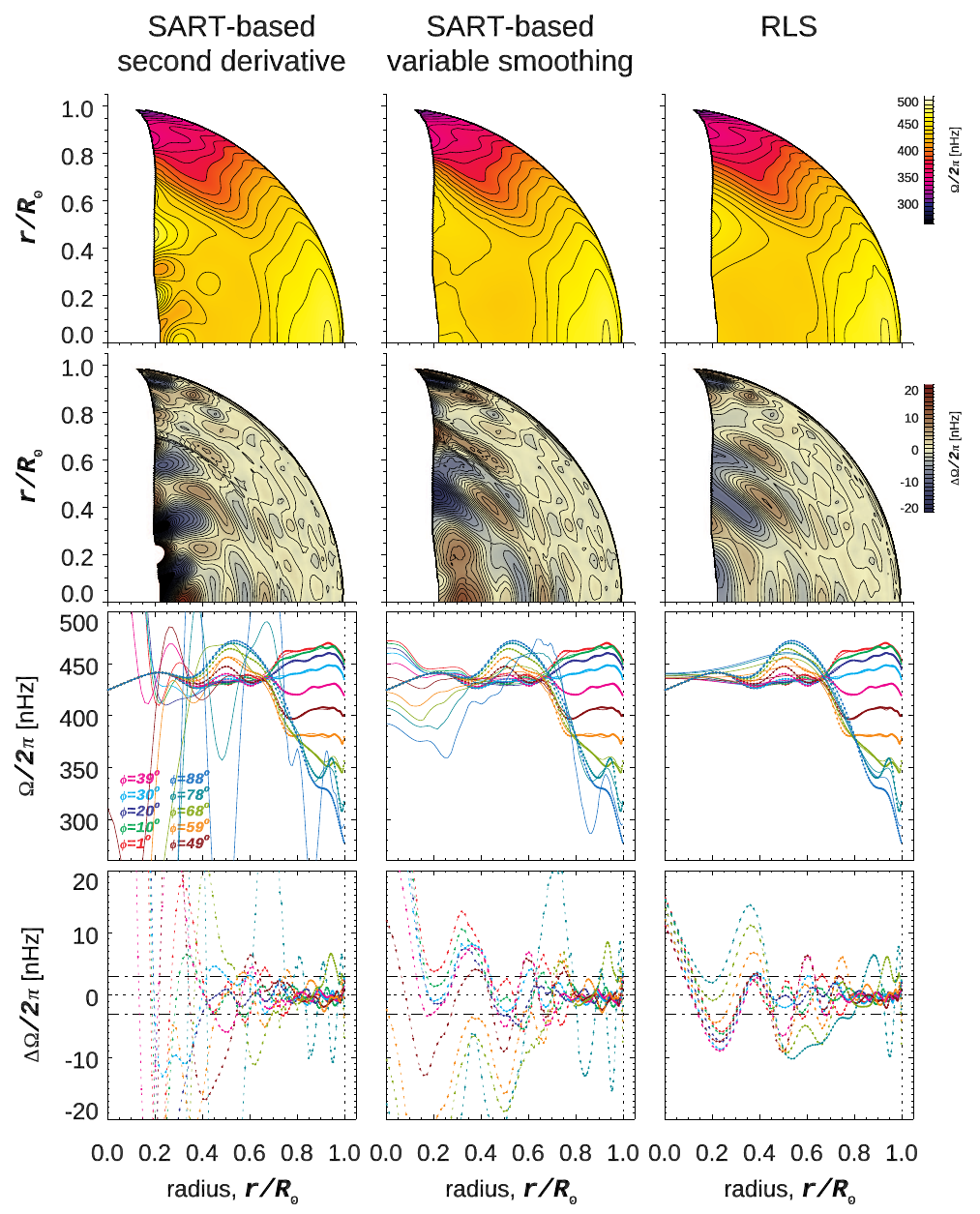} 
  \end{center}
  \caption{Inverted solution, based on artificial splittings with random noise,
    corresponding to model 1.
    The panels show the inferred solution and the difference between that
    inferred solution and the actual rotation model, in Cartesian coordinates in
    the top two rows or as cuts versus radius at a set of latitudes in the
    bottom two rows.
    The panels in the left column correspond to the SART-based inversion using a
    second derivative smoothing, the panels in the middle column correspond to the
    SART-based inversion using a variable smoothing used for regularization, while
    the panels in the rightmost column show the solution when using a conventional RLS
    inversion methodology.}\label{fig:mod1_error}
\end{figure}

\begin{figure}
  \begin{center}
    \includegraphics[width=0.975\textwidth]{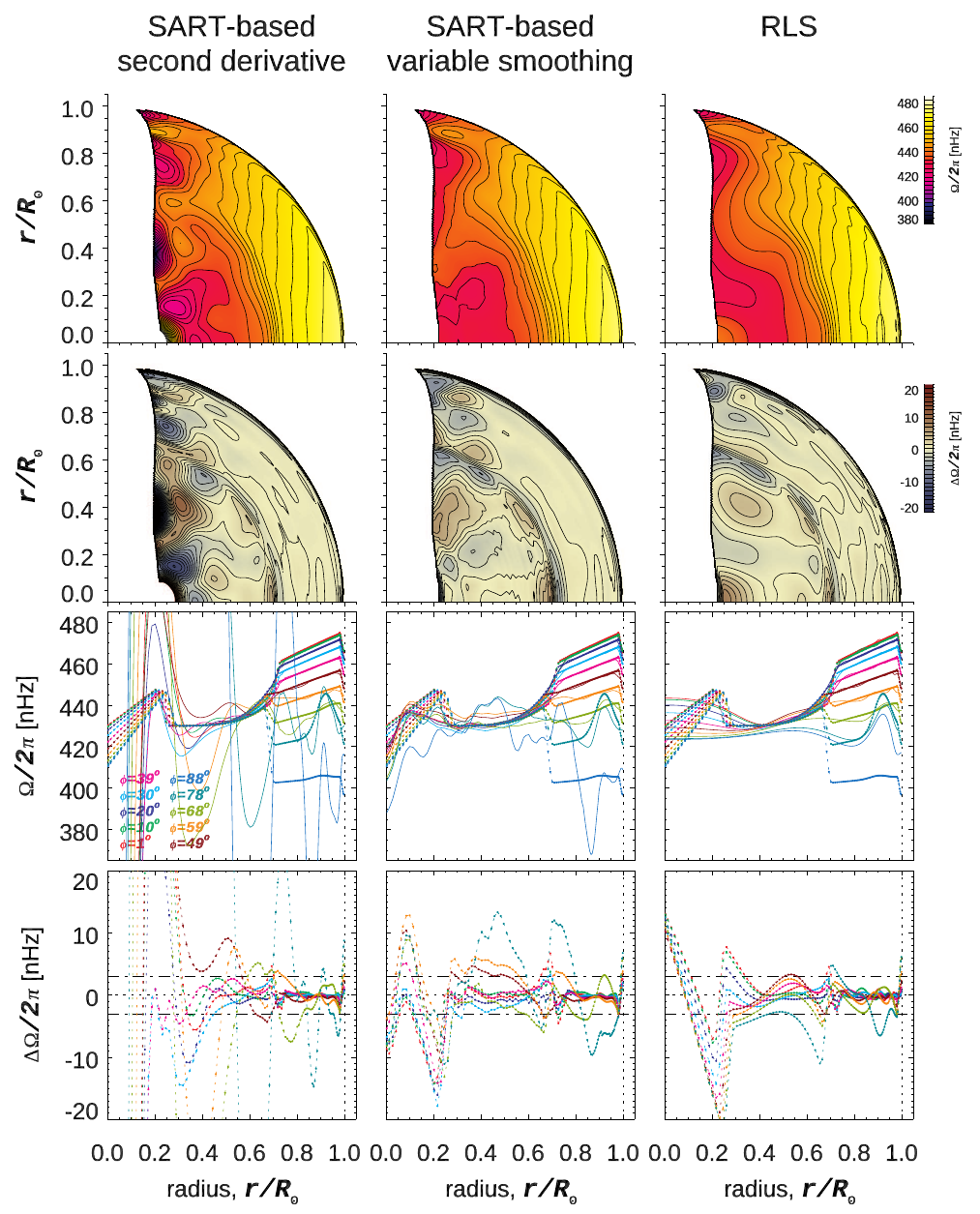} 
  \end{center}
  \caption{Inverted solution, based on artificial splittings with random noise,
    corresponding to model 2.
    The panels show the inferred solution and the difference between that
    inferred solution and the actual rotation model, in Cartesian coordinates in
    the top two rows or as cuts versus radius at a set of latitudes in
    the bottom two rows.
    The panels in the left column correspond to the SART-based inversion using a
    second derivative smoothing, the panels in the middle column correspond to the
    SART-based inversion using a variable smoothing used for regularization, while
    the panels in the rightmost column show the solution when using a conventional RLS
    inversion methodology.}\label{fig:mod2_error}
\end{figure}

Figures~\ref{fig:mod1_error} and \ref{fig:mod2_error} show the inferred
rotational profiles when inverting artificial splittings with random noise,
for each model respectively. As for the noiseless cases each figure presents
the inferred rotation rate using the SART-based inversion method when using
the same two different regularizations (a second derivative smoothing or a
variable smoothing) and the inferred profile when using a conventional RLS
inversion methodology. For these results, we apply a mask to hide the regions
where the inversion results are unreliable, namely where the
corresponding averaging kernels are not localized anywhere close to the target
location.

For each combination of methodology and regularization, the inferred solutions
recover the overall character of the underlying true rotation profile, but
with different levels of precision at different positions and not quite the
same departure from the true solution for each model.
Not surprisingly, the solutions are quite good at low latitudes in the
convection zone. Using a variable smoothing for regularization improves the
SART-based solutions in the radiative zone, since the smoothing increases with
depth and latitude.

One also notice that the SART-based solutions deviate more from the true
solution at the highest latitudes, compared to the RLS solutions, but the
specific nature of the solution, especially for model 2, namely, a rotation
rate constant on cylinders, is better recovered with the SART-based
method. Indeed the spurious oscillation seen in the RLS solution is absent in
the SART-based solutions and the high latitude jet of model 2 is more clearly
apparent, although somewhat displaced, for the SART-based solution with variable
smoothing.
The RMS of the differences with respect to the input model profile for the
inversion with random noise are tabulated in Table~\ref{tab:error}.

\begin{figure}
  \begin{center}
    \includegraphics[width=0.975\textwidth]{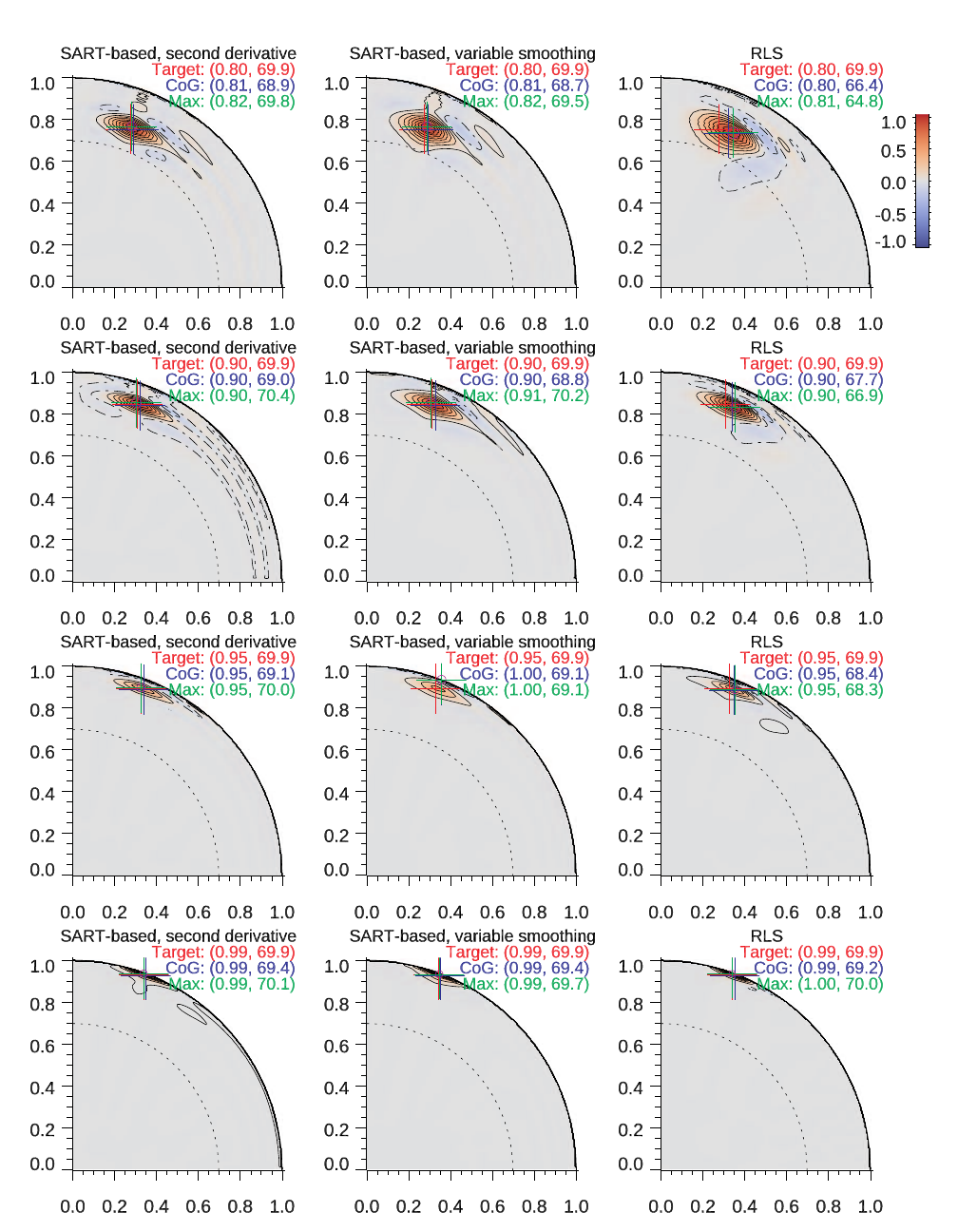} \\ 
  \end{center}
  \caption{Averaging kernels for solutions shown in Figures \ref{fig:mod1_error} and
    \ref{fig:mod2_error}, when using artificial splittings with random
    noise. 
    {Remember that in the presence of noise, for all other things equal, the averaging 
      kernels become wider since the regularization had to be increased to limit the noise 
      magnification. }
    The panels in each row show them at various depths near the surface
    ($r/R_\odot=0.8, 0.9, 0.95, 0.99$) at a high target latitude, namely
    $\phi=70^{\circ}$. 
    The panels on the left correspond to the SART-based method when using a
    second derivative smoothing, the panels in the middle to the SART-based
    method when using a variable smoothing for regularization, while the panels
    on the right correspond to using a RLS inversion
    methodology.
    The position of the target location, the center of gravity, and the maximum
    of the kernels are color coded, marked as crosses and labeled as $(r,
    \phi)$.}\label{fig:avgkern-error-l70s}
\end{figure}

\begin{figure}\begin{center}
    \includegraphics[width=0.975\textwidth]{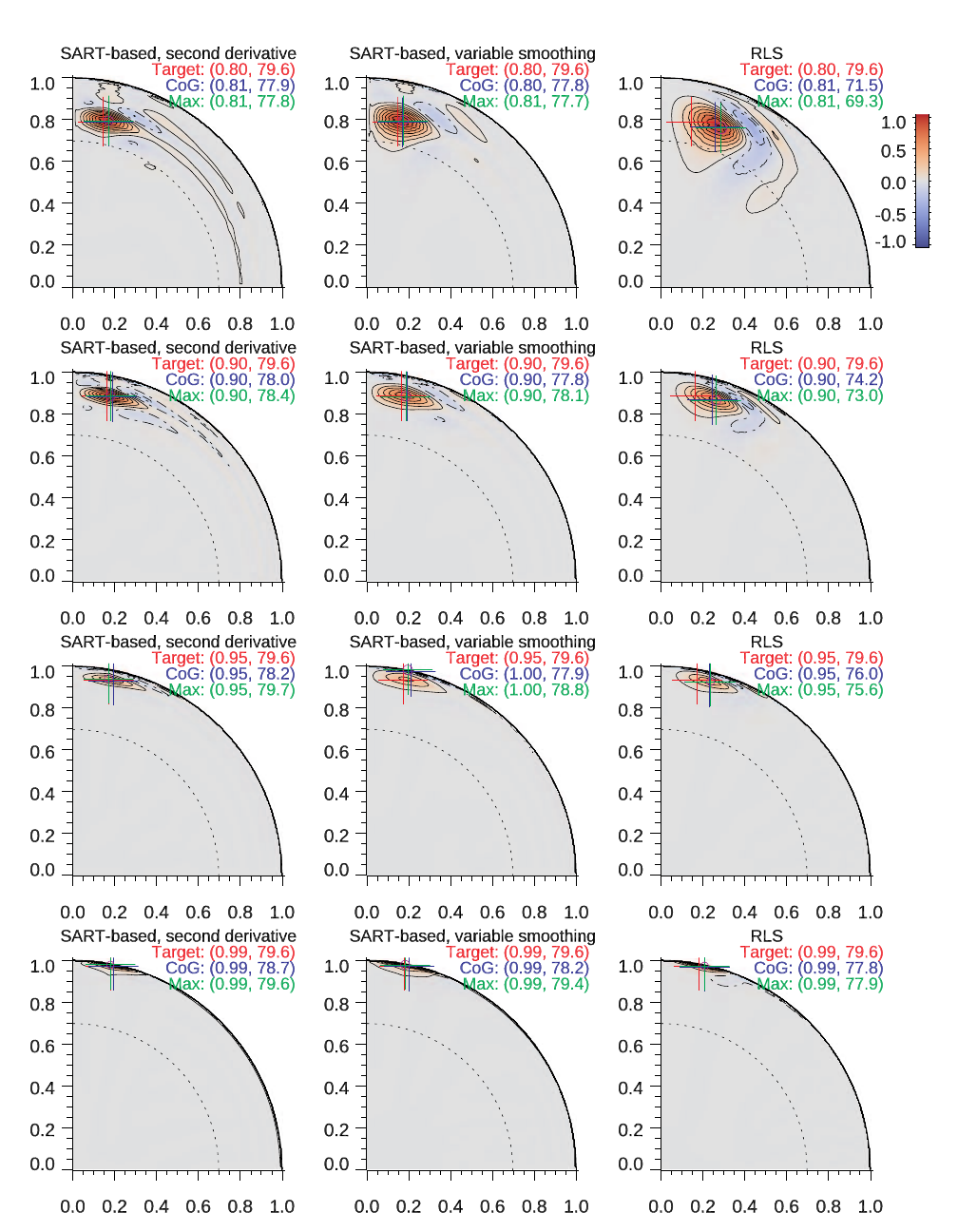} 
  \end{center}
  \caption{Averaging kernels for solutions shown in Figures \ref{fig:mod1_error} and
    \ref{fig:mod2_error}, when using artificial splittings with random
    noise. 
    %
    The panels in each row show them at various depths near the surface
    ($r/R_\odot=0.8, 0.9, 0.95, 0.99$) at an even higher target latitude, namely
    $\phi=80^{\circ}$.
    The panels on the left correspond to the SART-based method when using a
    second derivative smoothing, the panels in the middle to the SART-based
    method when using a variable smoothing for regularization, while the panels
    on the right correspond to using a RLS inversion
    methodology.
    The position of the target location, the center of gravity, and the maximum
    of the kernels are color coded, marked as crosses and labeled as $(r,
    \phi)$.}\label{fig:avgkern-error-l80s}
\end{figure}

\begin{figure}
  \begin{center}
    \includegraphics[width=0.975\textwidth]{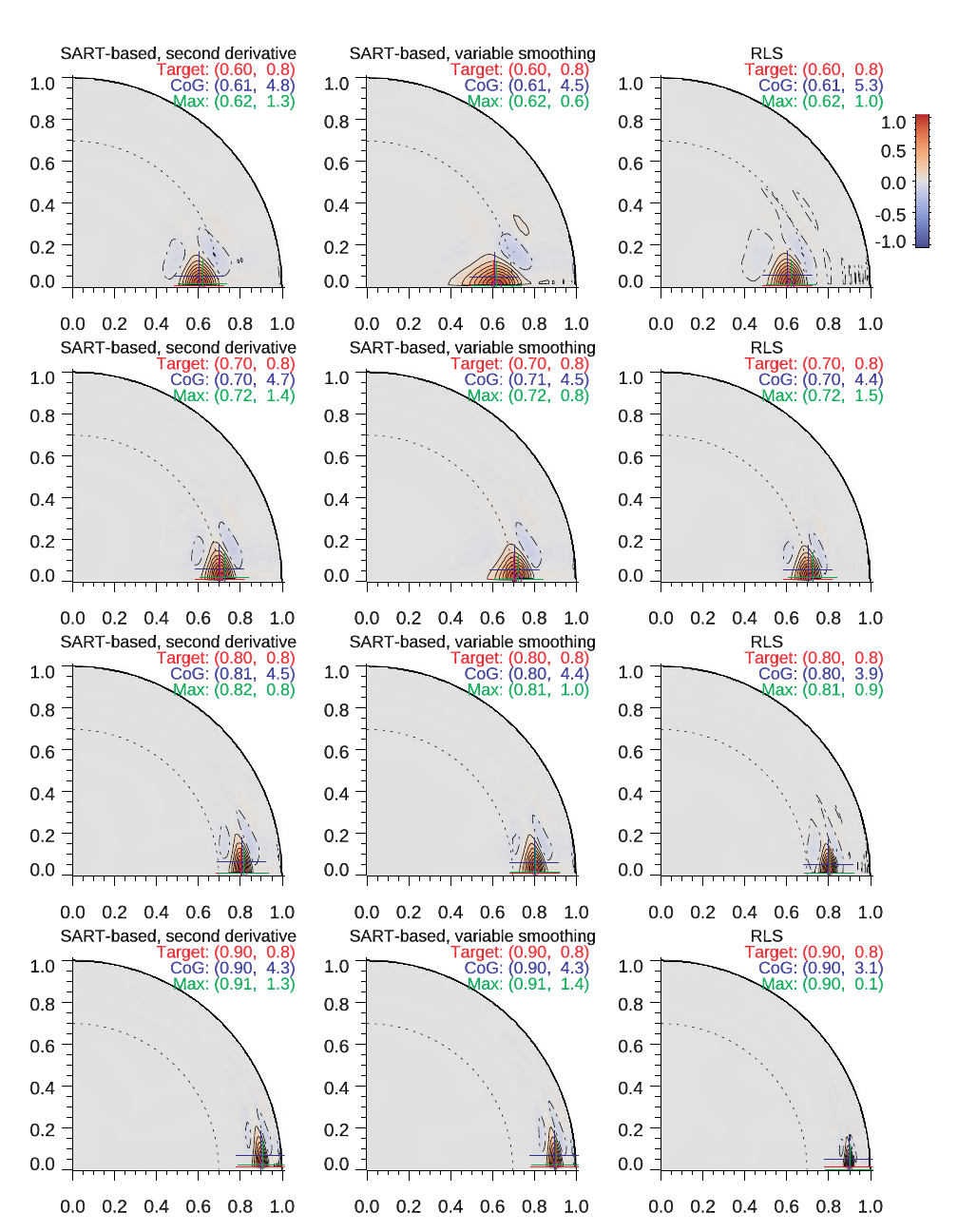} \\ 
  \end{center}
  \caption{Averaging kernels for solutions shown in Figures \ref{fig:mod1_error} and
    \ref{fig:mod2_error}, when using artificial splittings with random
    noise. 
    The panels in each row show them at various depths in the outer 40\% 
    ($r/R_\odot=0.6, 0.7, 0.8, 0.9$) at the equatorial target latitude, namely
    $\phi=0^{\circ}$. 
    The panels on the left correspond to the SART-based method when using a
    second derivative smoothing, the panels in the middle to the SART-based
    method when using a variable smoothing for regularization, while the panels
    on the right correspond to using a RLS inversion
    methodology.
    The position of the target location, the center of gravity, and the maximum
    of the kernels are color coded, marked as crosses and labeled as $(r,
    \phi)$.}\label{fig:avgkern-error-l00m}
\end{figure}

\begin{figure}\begin{center}
    \includegraphics[width=0.975\textwidth]{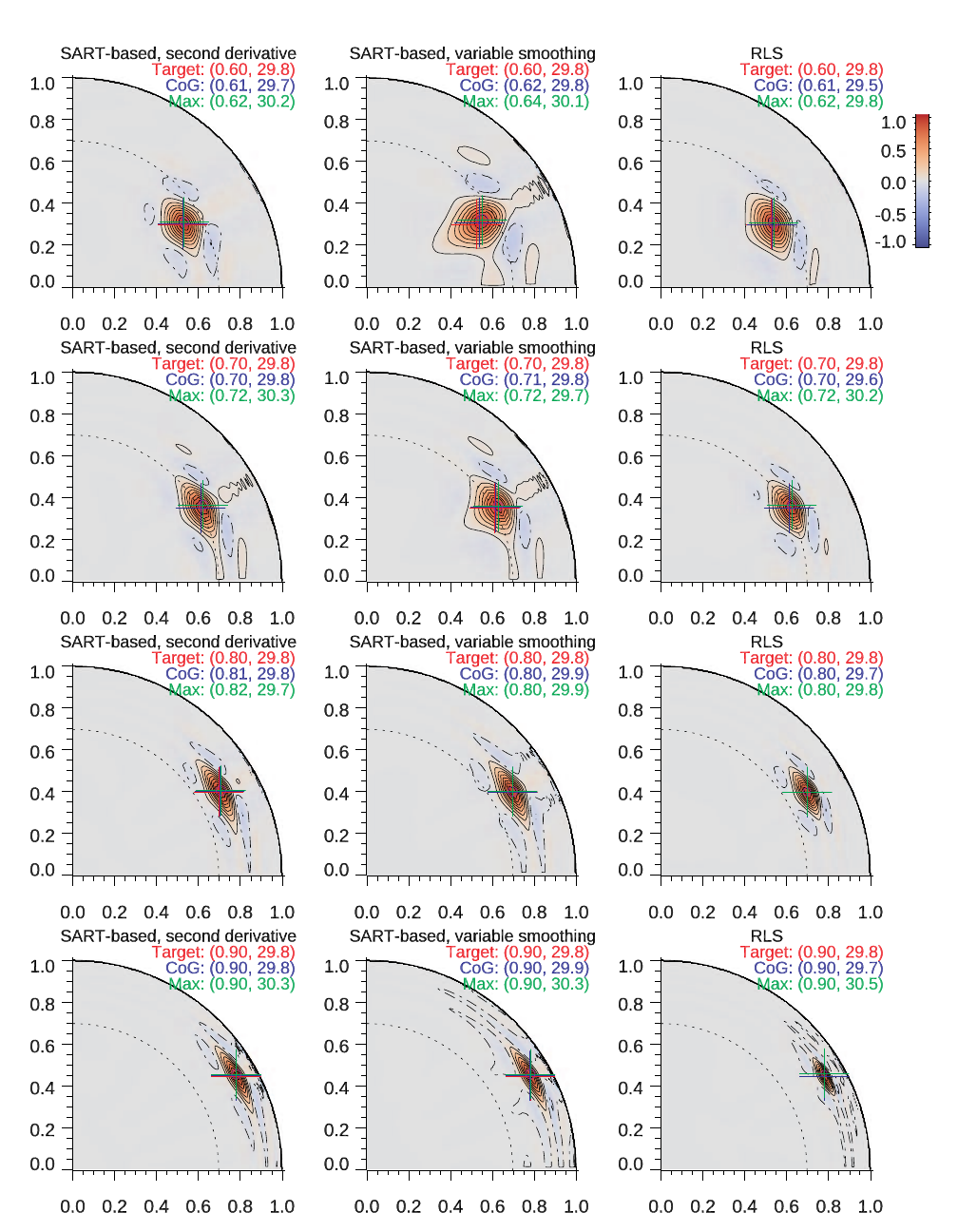} 
  \end{center}
  \caption{Averaging kernels for solutions shown in Figures \ref{fig:mod1_error} and
    \ref{fig:mod2_error}, when using artificial splittings with random
    noise. 
    The panels in each row show them at various depths in the outer 40\% 
    ($r/R_\odot=0.6, 0.7, 0.8, 0.9$) at a low target latitude, namely
    $\phi=30^{\circ}$. 
    The panels on the left correspond to the SART-based method when using a
    second derivative smoothing, the panels in the middle to the SART-based
    method when using a variable smoothing for regularization, while the panels
    on the right correspond to using a RLS inversion
    methodology.
    The position of the target location, the center of gravity, and the maximum
    of the kernels are color coded, marked as crosses and labeled as $(r,
    \phi)$.}\label{fig:avgkern-error-l30m}
\end{figure}

\begin{figure}
  \begin{center}
    \includegraphics[width=0.975\textwidth]{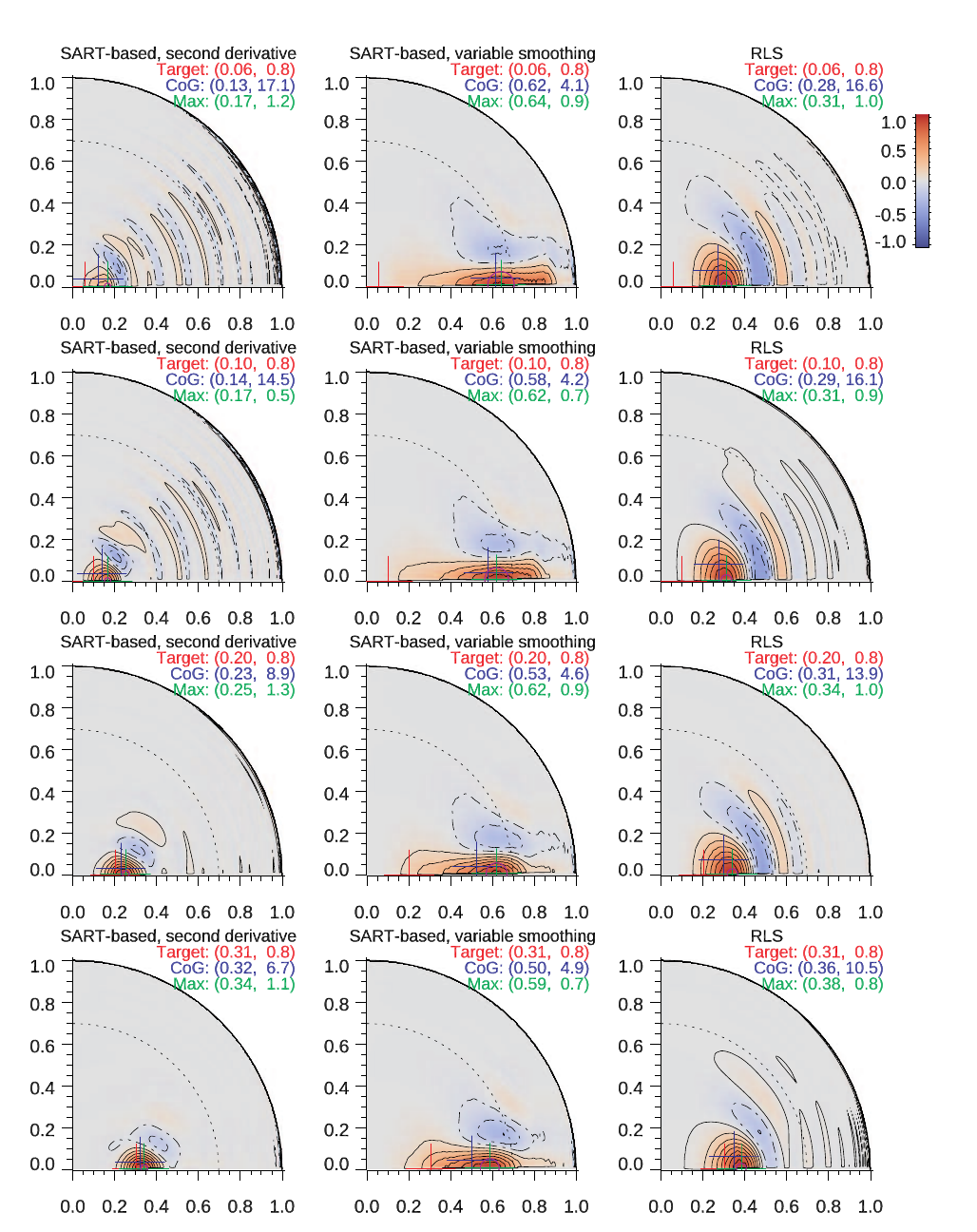} 
  \end{center}
  \caption{Averaging kernels for solutions shown in Figures \ref{fig:mod1_error} and
    \ref{fig:mod2_error}, when using artificial splittings with random
    noise. 
    The panels in each row show them at various depths in the inner 30\% 
    ($r/R_\odot=0.06, 0.1, 0.2, 0.3$) at the equatorial target latitude, namely
    $\phi=0^{\circ}$. 
    The panels on the left correspond to the SART-based method when using a
    second derivative smoothing, the panels in the middle to the SART-based
    method when using a variable smoothing for regularization, while the panels
    on the right correspond to using a RLS inversion
    methodology.
    The position of the target location, the center of gravity, and the maximum
    of the kernels are color coded, marked as crosses and labeled as $(r,
    \phi)$.}\label{fig:avgkern-error-l00c}
\end{figure}

\begin{figure}
  \begin{center}
    \includegraphics[width=0.975\textwidth]{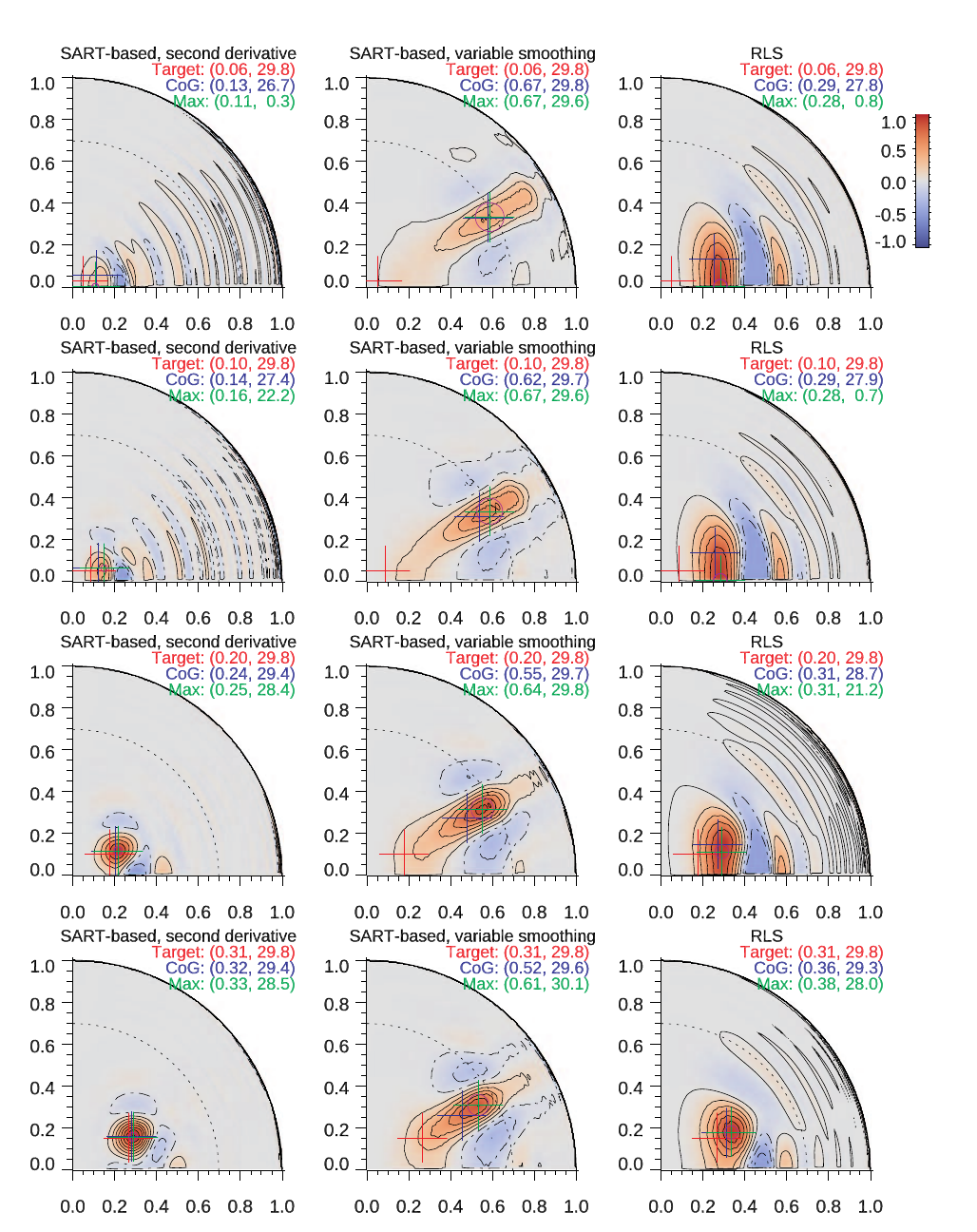} 
  \end{center}
  \caption{Averaging kernels for solutions shown in Figures \ref{fig:mod1_error} and
    \ref{fig:mod2_error}, when using artificial splittings with random
    noise. 
    The panels in each row show them at various depths in inner 30\% 
    ($r/R_\odot=0.06, 0.1, 0.2, 0.3$) at a low target latitude, namely
    $\phi=30^{\circ}$. 
    The panels on the left correspond to the SART-based method when using a
    second derivative smoothing, the panels in the middle to the SART-based
    method when using a variable smoothing for regularization, while the panels
    on the right correspond to using a RLS inversion
    methodology.
    The position of the target location, the center of gravity, and the maximum
    of the kernels are color coded, marked as crosses and labeled as $(r,
    \phi)$.}\label{fig:avgkern-error-l30c}
\end{figure}

The corresponding averaging kernels, for the same selected target locations as
for the noiseless cases, are shown in Figures~\ref{fig:avgkern-error-l70s} to
\ref{fig:avgkern-error-l30c}.
Comparing these to the noiseless cases shows overall wider kernels, since the
regularization is increased to dampen the noise. They remain well-localized
for the same target locations and vice versa. When comparing these with
respect to the inversion and smoothing methodology, the averaging kernels show
small and subtle differences when they are well-localized and wild
differences when they are not well-localized.

\subsection{Effect of Regularization}

\begin{figure}
  \begin{center}
    \includegraphics[width=0.975\textwidth]{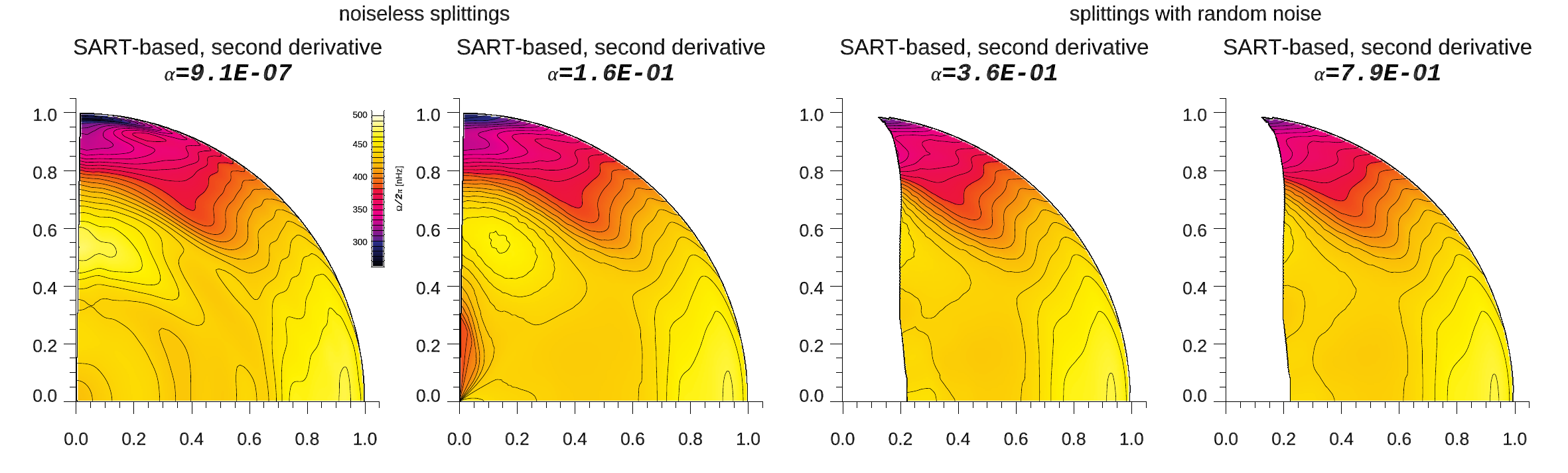} \\ 
    \includegraphics[width=0.975\textwidth]{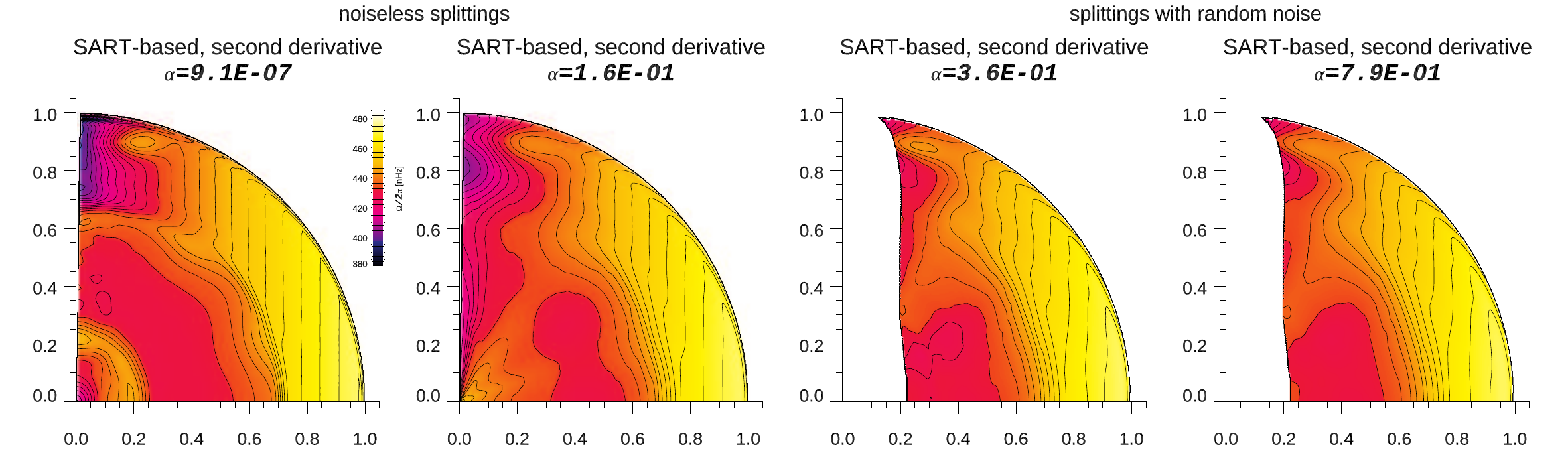} \\ 
  \end{center}
  \caption{Effect of adjusting the regularization factor, $\alpha$, when using the
    SART-based inversion methodology and the variable smoothing for
    regularization for either model (top and bottom panels), for both noiseless and noisy
    splittings for two values of $\alpha$ (left to right panels).}\label{fig:regularization}
\end{figure}

To further illustrate the SART-based method, Figure \ref{fig:regularization}
shows the effect of adjusting the regularization factor, $\alpha$. The figure
shows inferred solutions for both models, when using noiseless splittings and
when using splittings with random noise, for two values of $\alpha$ and the
variable smoothness regularization. In all cases, decreasing $\alpha$
increases the smoothing and thus washes out some of the underlying profile
features in exchange for reducing the noise amplification.

\subsection{Inversion of Artificial Data when Using a Non-Uniform Initial
  Guess}

\begin{figure}
  \begin{center}
    \includegraphics[width=0.975\textwidth]{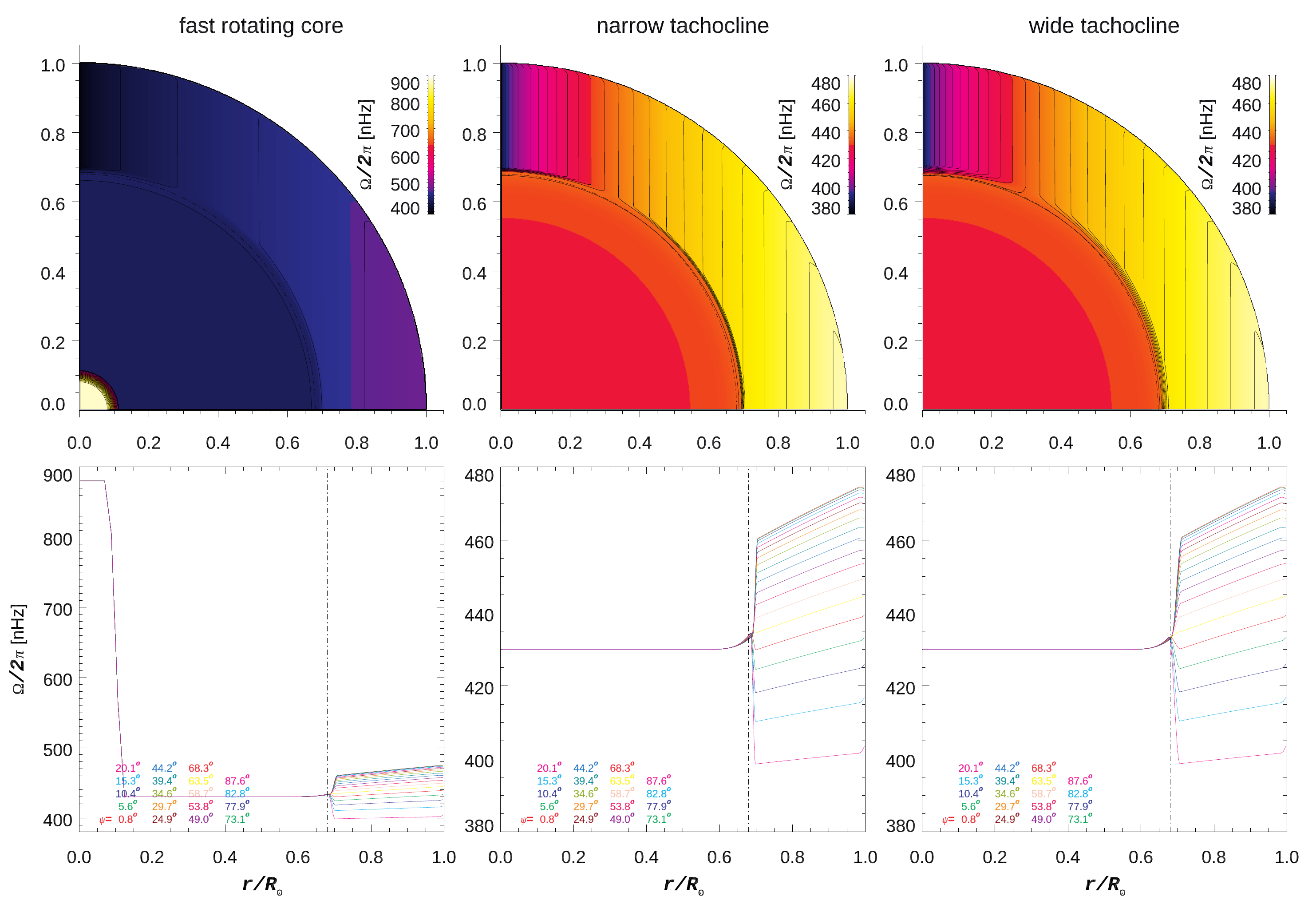} 
  \end{center}
  \caption{Additional rotation models used to compute artificial splittings.
    From left to right: models with (i) a rapidly rotating core, (ii) a narrow
    tachocline. and (iii) a wider tachocline.
    The top panels show them in Cartesian
    coordinates, while the bottom panels show cuts versus radius
    at selected latitudes.}\label{fig:addl-models}
\end{figure}

One feature of our SART-based inversion methodology is the ability to start
iterating from an arbitrary initial guess, one that for instance can be based
on prior knowledge. To test this feature, we used several simplified variants
of model 2, where we changed the width of the tachocline and removed the other
``{features}''. We also, in one case, add a rapidly rotating
core.
Figure \ref{fig:addl-models} shows some of these models, namely the one with a
fast rotating core and two models with a slightly different tachocline width:
a narrow one and a wider one.

\begin{figure}
  \begin{center}
    \includegraphics[width=0.975\textwidth]{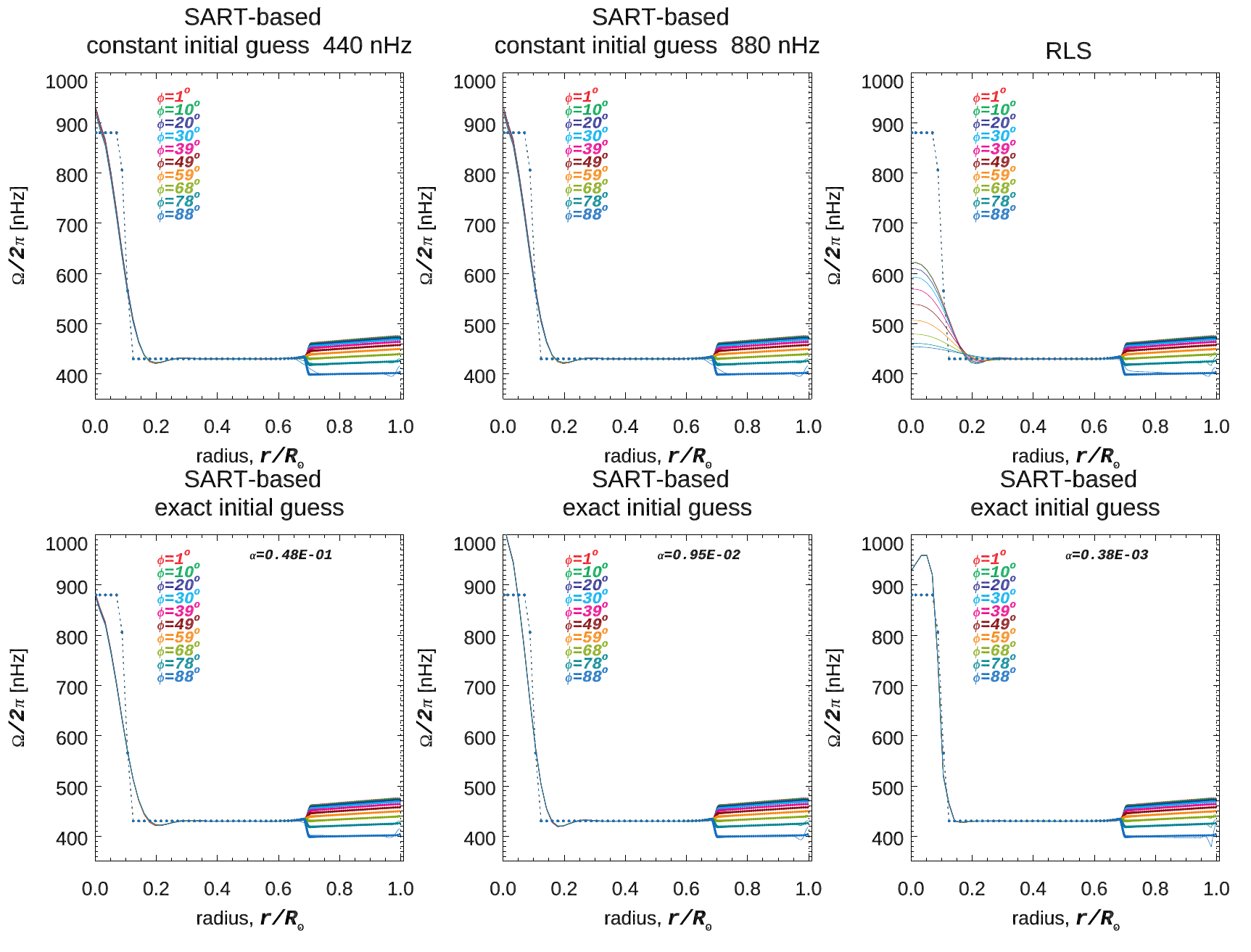} 
  \end{center}
  \caption{Inverted profiles when using artificial noiseless splittings
    corresponding to a rapidly rotating core. The solution for the RLS inversion
    {(upper right panel)} is compared to the SART-based inversion using the
    second derivative smoothing, a constant initial guess for two different
    values {(upper left and middle panels)}, and an initial guess that is the
    known underlying profile for three different regularization values {(lower
      panels)}.}\label{fig:rrc}
\end{figure}

Profiles obtained when inverting artificial noiseless splittings computed with
a rapidly rotating core profile are shown in Figure \ref{fig:rrc}, when using
either the RLS or the SART-based method. For the SART-based method the figure 
shows inferences obtained when using (i) two very different values of a constant 
initial guess, or (ii) the known rotation profile as initial guess and three different
values of the regularization factor, $\alpha$. 
While the RLS inversion hints at a rapidly rotating core,
the rotation at the center remains well underestimated. In contrast, even with
only $p$-modes, the SART-based inversions recover the fast rotating core rate,
but not its precise profile. Using the known rotation profile as an initial
guess does a better job at some value of the regularization factor, but
overestimates the rate at the center. The SART-based method, even with the
correct initial guess and noiseless splittings, is no match to the marginal
sensitivity of the $p$-modes to the rotation rate of the core.

\begin{figure}
  \begin{center}
    \includegraphics[width=0.975\textwidth]{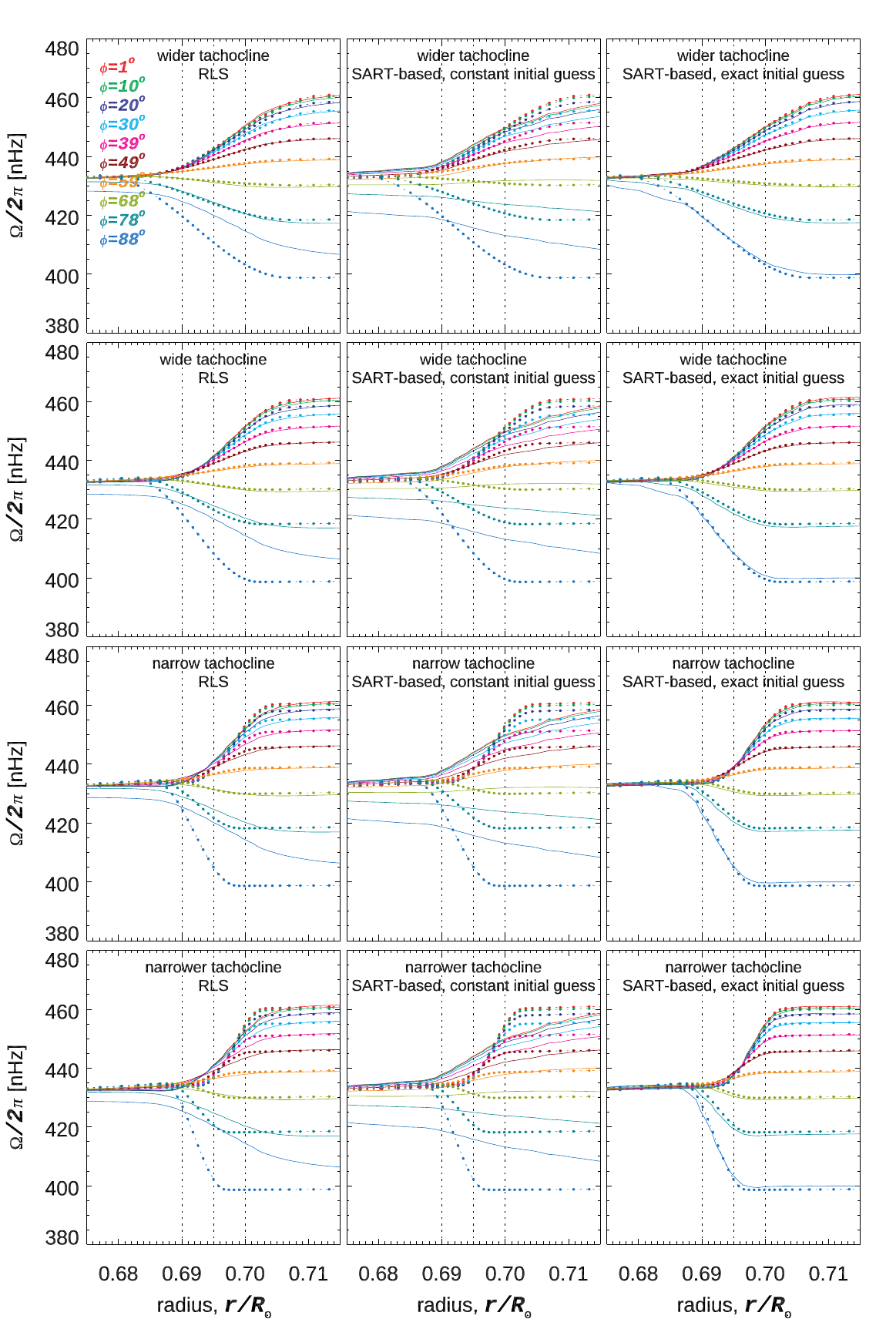} 
  \end{center}
  \caption{Inverted profiles when using artificial noiseless splittings
    corresponding to four different thicknesses of the tachocline (top to bottom
    panels). The solution for the RLS inversion is compared to the SART-based
    inversion using a constant initial guess and an initial guess that is the
    known underlying profile (left to right panels). Fiducial lines are added
    only to guide the eye since only a small fraction of the radial range is
    shown.}\label{fig:tacho}
\end{figure}

Figure \ref{fig:tacho} shows rotation profiles obtained when inverting artificial
noiseless splittings computed for models with various thicknesses of the
tachocline. The figure compares the RLS method to the SART-based method using
either a constant as the initial guess or the known rotation profile as the initial
guess. In the noiseless case, the SART-based method does recover perfectly the
tachocline profiles if and when the initial guess is the known rotation
profile. In fact, the SART-based method does not do as well as the RLS method
when using a constant as an initial guess, especially for very narrow
tachoclines, but outperforms it when using the right initial guess.

\begin{figure}
  \begin{center}
    \includegraphics[width=0.975\textwidth]{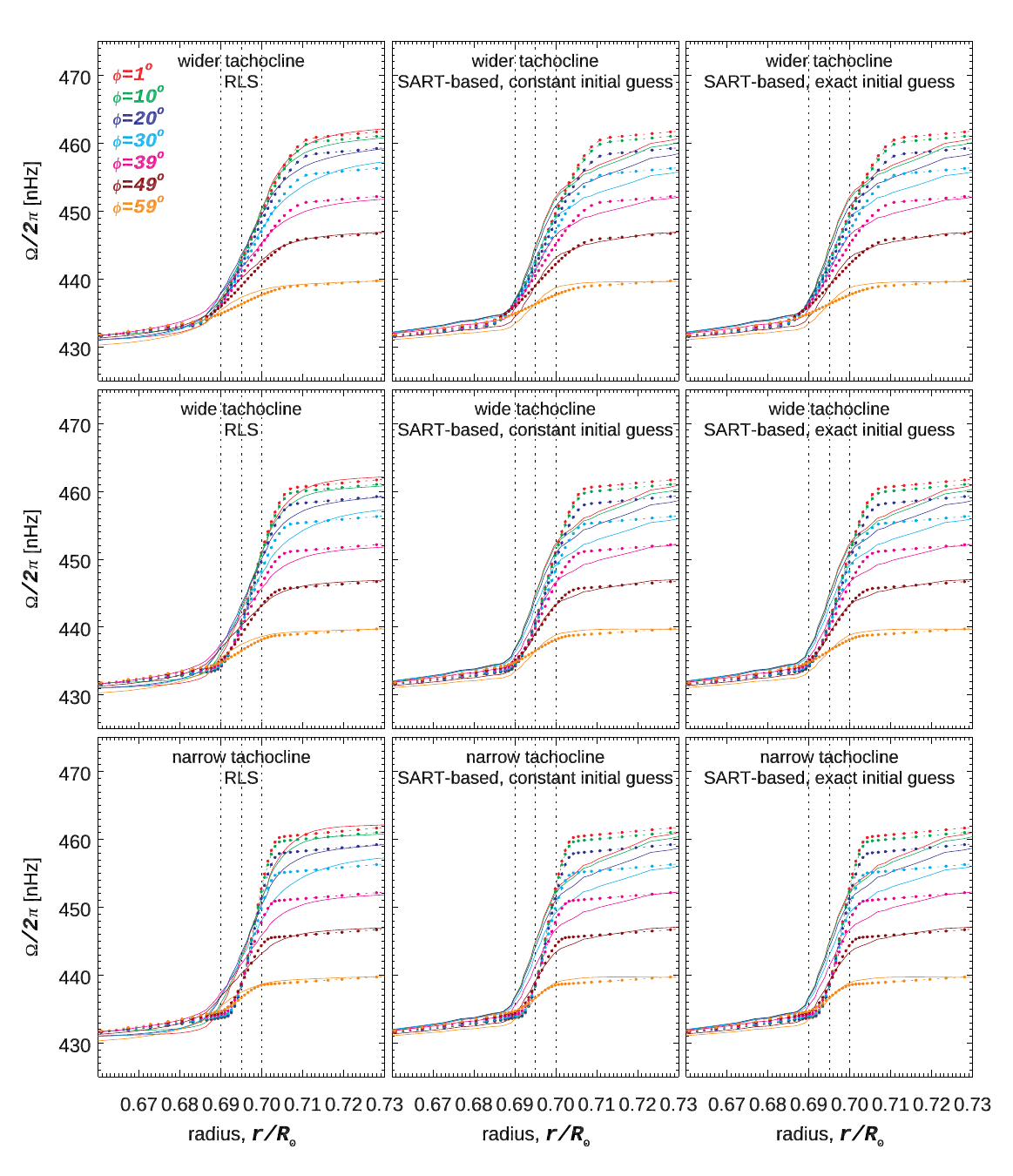} 
  \end{center}
  \caption{Inverted profiles when using artificial splittings with little random
    noise corresponding to three different thicknesses of the tachocline (top to
    bottom panels). the solution for the RLS inversion is compared to the
    SART-based inversion using a constant initial guess and an initial guess
    that is the known underlying profile (left to right panels). Fiducial lines
    are added only to guide the eye since only a small fraction of the radial
    range is shown and for a limited range in latitudes.}\label{fig:tacho+nf}
\end{figure}

\begin{figure}
  \begin{center}
    \includegraphics[width=0.975\textwidth]{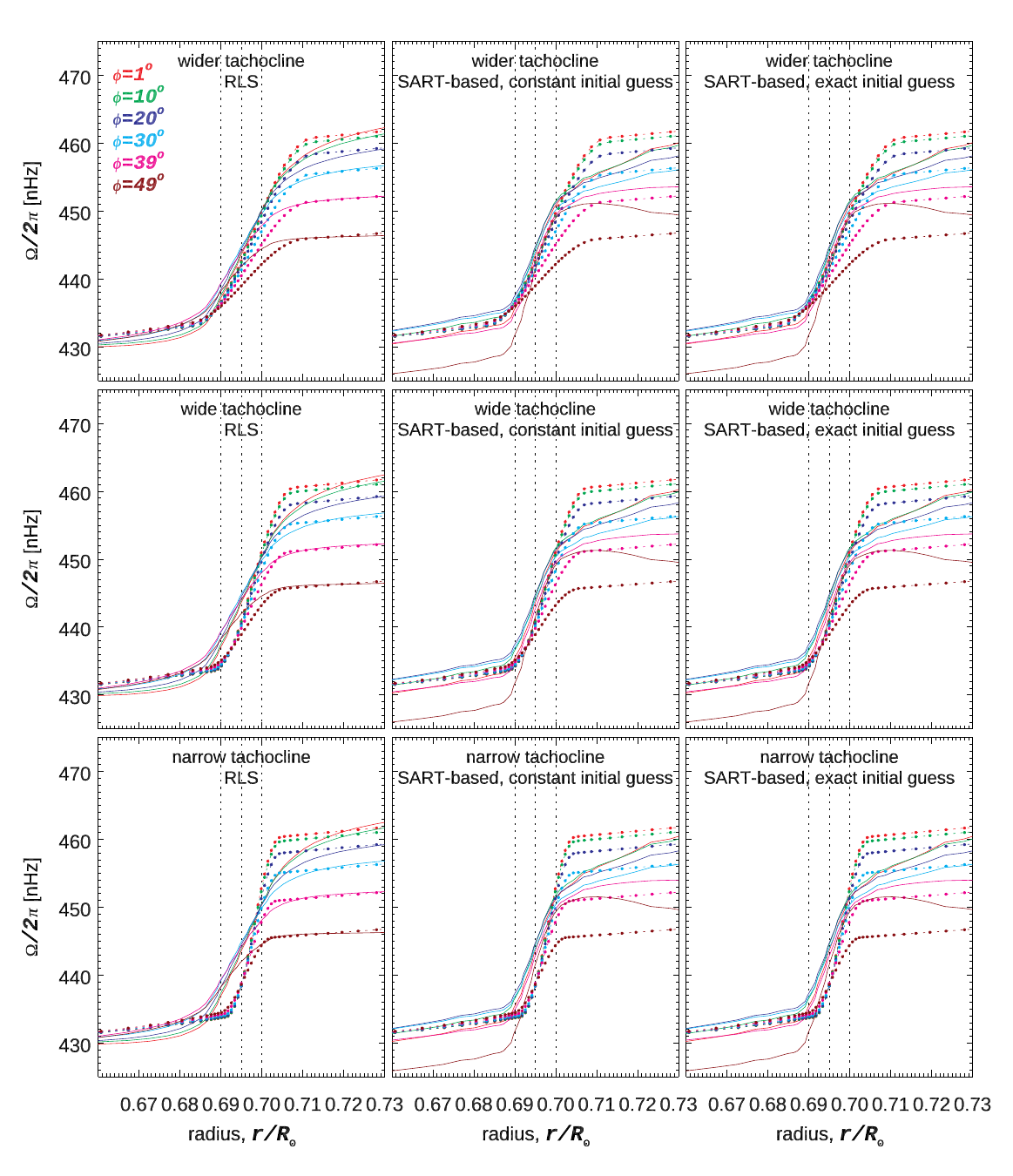} 
  \end{center}
  \caption{Inverted profiles when using artificial splittings with random noise
    corresponding to three different thicknesses of the tachocline (top to bottom
    panels). The solution for the RLS inversion is compared to the SART-based
    inversion using a constant initial guess and an initial guess that is the
    known underlying profile (left to right panels). {This figure shows that the more 
      narrow tachocline models are not well recovered in all cases, but also how the SART-based 
      inferences present other systematic errors near the tachocline at higher latitudes.}  
    Fiducial lines are added only to guide the eye since only a small fraction of the radial range is
    shown and for a limited range in latitudes.}\label{fig:tacho+nf10}
\end{figure}

Figures \ref{fig:tacho+nf} and \ref{fig:tacho+nf10} show rotation profiles
obtained when inverting artificial splittings with just a little noise or
random noise corresponding to the observations and computed for some of the
same models with various thicknesses of the tachocline. Once we add some random
noise to the data, the benefit of using the right initial guess is
unfortunately washed out, as our current implementation shows a somewhat
limited immunity to noise. We plan on further investigating how to increase
this noise immunity, since our SART-based method gives us a lot more
flexibility in the implementation of the regularization and hence the
smoothing.

\section{Conclusions} 

We have described the new iterative rotation inversion methodology we have
implemented based on the SART algorithm and have compared its performances when using
artificial rotational splittings to a classical RLS inversion
method. Although iterative, our implementation allows us to compute directly
both the formal uncertainties on the solution, via standard error propagation,
and the averaging kernels without having recourse to some form of
Monte-Carlo simulation. The regularization of the inverse problem via local
smoothing has been implemented in a very flexible way, using a generalized
formalism.
In various noiseless cases as well as realistic cases with random noise, this
new technique outperforms the RLS, in precision, scope and resolution. While
its immunity to noise is lower, our flexible regularization allows selective
damping of this noise magnification.

The resulting averaging kernels are only slightly dissimilar to those of an
RLS method where they are well-localized. In contrast, they differ by a lot
where the solution is not well constrained because of poorly-localized
averaging kernels.

In our current implementation, the use of some apriori information, like the
sharpness of the tachocline, works remarkably well in noiseless cases, but is
washed out in the presence of noise. This is an area that calls for further
work, since such an iterative method could be used to improve our resolution
of the tachocline. Indeed, a classical RLS combined with some form of
deconvolution could be used to provide a very good initial guess to our
iterative technique, once we improve the SART-based methodology noise immunity
near the tachocline.

While we do not anticipate that this method will relegate the classical RLS or SOLA
methods to the dustbin, using more that one technique to invert real observations
can only improve the robustness of our inferences. Comparing and contrasting results 
obtained when using different inversion techniques allow us to better
understand where these might be biased and/or unreliable, especially as we push our
techniques to maximize the diagnostic potential of our observations.

\begin{acknowledgements}

%
\begin{fundinginformation}
This work was partially supported by NASA grants 80\-NSSC22K0516 and \\
NNH18ZDA001N--DRIVE to SGK and by the Spanish AEI programs
PID2019--107187GB--I00 (STrESS), PID2019--104571RA--I00 (COMPACT),
PID2022--139159NB--I00 (Volca-Motion) \\
and PID2022--140483NB--C21 (HARMONI) to AED.
\end{fundinginformation}
\end{acknowledgements}

\newpage

\begin{table}[h!]
~\\*[0.15em]
\begin{center}\begin{turn}{90}
  \begin{tabular}{r|rrrr|rrrr|rrrr|}
             & \multicolumn{4}{c|}{SART-based}
             & \multicolumn{4}{c|}{SART-based}
             & \multicolumn{4}{c|}{RLS} \\
             & \multicolumn{4}{c|}{second derivative}
             & \multicolumn{4}{c|}{variable smoothing}
             & \multicolumn{4}{c|}{~} \\ \hline
    $r/R_{\odot}$
    & [0-0.3] & [0.3-0.7] & [0.7-1] & [0-1]
    & [0-0.3] & [0.3-0.7] & [0.7-1] & [0-1]
    & [0-0.3] & [0.3-0.7] & [0.7-1] & [0-1] \\ \hline
    lat. & \multicolumn{12}{c|}{model 1} \\\hline
 0.8 &  2.3 &  0.2 &  0.1 &  1.0 &  1.2 &  0.2 &  0.1 &  0.5 &  6.0 &  0.5 &  0.1 &  2.6 \\
10.5 &  2.3 &  0.1 &  0.1 &  1.0 &  1.2 &  0.1 &  0.2 &  0.5 &  6.2 &  0.3 &  0.1 &  2.7 \\
20.1 &  2.4 &  0.1 &  0.1 &  1.0 &  1.3 &  0.1 &  0.1 &  0.6 &  7.1 &  0.3 &  0.1 &  3.0 \\
29.8 &  2.4 &  0.1 &  0.2 &  1.0 &  1.4 &  0.1 &  0.2 &  0.6 &  8.0 &  0.2 &  0.2 &  3.4 \\
39.4 &  2.5 &  0.2 &  0.2 &  1.1 &  1.7 &  0.2 &  0.2 &  0.7 &  8.2 &  0.3 &  0.2 &  3.5 \\
49.0 &  2.6 &  0.2 &  0.5 &  1.2 &  2.0 &  0.2 &  0.5 &  0.9 &  7.6 &  0.7 &  0.5 &  3.3 \\
58.7 &  2.8 &  0.4 &  0.8 &  1.3 &  2.3 &  0.3 &  0.9 &  1.2 &  6.6 &  1.1 &  0.8 &  2.9 \\
68.3 &  2.9 &  0.4 &  1.5 &  1.7 &  2.6 &  0.4 &  1.4 &  1.5 &  5.7 &  2.0 &  1.4 &  2.9 \\
78.0 &  3.0 &  0.8 &  2.6 &  2.3 &  2.6 &  0.9 &  2.6 &  2.3 &  5.0 &  3.4 &  2.3 &  3.3 \\ \hline
    lat. & \multicolumn{12}{c|}{model 2} \\\hline

 0.8 &  8.3 &  0.4 &  0.2 &  3.5 &  7.1 &  0.3 &  0.2 &  3.0 & 10.4 &  0.3 &  0.2 &  4.4 \\
10.5 &  8.1 &  0.2 &  0.1 &  3.4 &  6.8 &  0.2 &  0.1 &  2.9 & 10.8 &  0.3 &  0.2 &  4.6 \\
20.1 &  7.4 &  0.2 &  0.1 &  3.1 &  6.1 &  0.2 &  0.1 &  2.6 & 11.5 &  0.2 &  0.1 &  4.9 \\
29.8 &  6.3 &  0.2 &  0.2 &  2.7 &  5.1 &  0.2 &  0.1 &  2.2 & 12.5 &  0.3 &  0.2 &  5.3 \\
39.4 &  5.1 &  0.2 &  0.3 &  2.2 &  3.9 &  0.2 &  0.3 &  1.7 & 13.5 &  0.3 &  0.3 &  5.7 \\
49.0 &  3.9 &  0.3 &  0.5 &  1.7 &  2.9 &  0.3 &  0.5 &  1.3 & 14.3 &  0.4 &  0.5 &  6.1 \\
58.7 &  2.9 &  0.4 &  0.8 &  1.4 &  2.4 &  0.4 &  0.9 &  1.2 & 14.8 &  0.7 &  0.8 &  6.3 \\
68.3 &  2.4 &  0.5 &  1.5 &  1.5 &  2.6 &  0.5 &  1.5 &  1.6 & 15.1 &  0.8 &  1.5 &  6.5 \\
78.0 &  2.7 &  1.1 &  3.0 &  2.5 &  3.4 &  1.0 &  3.0 &  2.7 & 15.3 &  2.3 &  3.5 &  7.1 \\ \hline
  \end{tabular}
\end{turn}\end{center}
\caption{RMS of the differences with respect to the input model profile for
    cuts at various latitudes when \\ inverting noiseless artificial
    splittings.}\label{tab:noerr}
\end{table}

\begin{table}[h!]
~\\*[.15em]
\begin{center}\begin{turn}{90}
  \begin{tabular}{r|rrrr|rrrr|rrrr|}
             & \multicolumn{4}{c|}{SART-based}
             & \multicolumn{4}{c|}{SART-based}
             & \multicolumn{4}{c|}{RLS} \\
             & \multicolumn{4}{c|}{second derivative}
             & \multicolumn{4}{c|}{variable smoothing}
             & \multicolumn{4}{c|}{~} \\ \hline
    $r/R_{\odot}$
    & [0-0.3] & [0.3-0.7] & [0.7-1] & [0-1]
    & [0-0.3] & [0.3-0.7] & [0.7-1] & [0-1]
    & [0-0.3] & [0.3-0.7] & [0.7-1] & [0-1] \\ \hline
    lat. & \multicolumn{12}{c|}{model 1} \\\hline
 0.8 & 483.1 &  6.9 &  1.1 & 205.6 & 26.1 &  6.3 &  1.5 & 11.7 &  7.7 &  2.1 &  1.1 &  3.6 \\
10.5 & 474.8 &  2.8 &  1.1 & 202.0 & 24.0 &  4.8 &  1.4 & 10.6 &  7.6 &  1.5 &  0.7 &  3.4 \\
20.1 & 431.6 &  5.5 &  0.9 & 183.7 & 21.2 &  3.8 &  0.9 &  9.3 &  7.3 &  1.4 &  0.7 &  3.2 \\
29.8 & 315.8 &  4.7 &  1.0 & 134.4 & 18.5 &  4.4 &  1.1 &  8.2 &  7.1 &  2.1 &  0.8 &  3.3 \\
39.4 & 127.0 &  6.5 &  0.8 &  54.1 & 12.5 &  5.0 &  1.1 &  6.0 &  6.9 &  3.7 &  0.7 &  3.6 \\
49.0 & 138.4 & 10.8 &  1.1 &  59.2 &  8.3 &  4.5 &  1.6 &  4.4 &  6.2 &  4.4 &  1.2 &  3.6 \\
58.7 & 354.4 &  9.3 &  1.7 & 150.9 & 19.1 &  8.2 &  2.2 &  9.3 &  5.6 &  4.9 &  2.0 &  3.8 \\
68.3 & 544.6 & 21.1 &  3.2 & 232.0 & 29.2 & 11.5 &  4.1 & 14.2 &  6.4 &  7.4 &  3.0 &  5.3 \\
78.0 & 703.0 & 61.6 & 10.4 & 301.0 & 40.8 & 14.9 & 10.2 & 20.5 &  7.9 &  9.3 &  5.1 &  7.1 \\ \hline
    lat. & \multicolumn{12}{c|}{model 2} \\\hline
 0.8 & 599.2 &   2.9 &  0.7 & 255.0 &  7.5 &  2.0 &  1.3 &  3.5 &  7.0 &  1.9 &  0.8 &  3.2 \\
10.5 & 600.1 &   1.8 &  0.5 & 255.3 &  7.2 &  1.0 &  0.9 &  3.2 &  6.9 &  1.3 &  0.6 &  3.0 \\
20.1 & 609.9 &   4.3 &  0.4 & 259.5 &  7.4 &  1.0 &  0.8 &  3.3 &  6.7 &  0.8 &  0.4 &  2.9 \\
29.8 & 620.4 &   5.9 &  0.4 & 264.0 &  8.0 &  2.0 &  0.8 &  3.6 &  7.4 &  1.2 &  0.6 &  3.3 \\
39.4 & 608.7 &   1.7 &  0.6 & 259.0 &  6.5 &  1.0 &  1.0 &  2.9 &  8.7 &  1.3 &  0.7 &  3.8 \\
49.0 & 557.6 &   6.1 &  0.9 & 237.3 &  6.2 &  2.9 &  1.3 &  3.2 &  9.7 &  2.1 &  0.8 &  4.3 \\
58.7 & 484.6 &  14.3 &  1.7 & 206.3 &  6.5 &  4.0 &  1.8 &  3.7 & 10.6 &  2.2 &  1.4 &  4.8 \\
68.3 & 417.9 &  30.1 &  3.4 & 178.5 &  6.8 &  2.9 &  2.6 &  3.8 & 11.3 &  2.5 &  2.0 &  5.2 \\
78.0 & 408.3 & 111.4 &  9.4 & 183.7 &  6.2 &  7.2 &  6.9 &  6.9 & 11.8 &  5.4 &  3.6 &  6.3 \\ \hline
  \end{tabular}
\end{turn}\end{center}
  \caption{RMS of the differences with respect to the input model profile for
    cuts at various latitudes when \\ inverting artificial splittings with random
    noise.}\label{tab:error}
\end{table}

\clearpage
%

%
\end{article} 
\end{document}